\documentclass[article, onecolumn, showpacs, 11pt]{revtex4-1}
\usepackage{color}
\usepackage{amsmath}
\usepackage{amssymb}
\usepackage{graphicx}
\usepackage{hyperref}
\hypersetup{
    colorlinks,%
    citecolor=blue,%
    filecolor=blue,%
    linkcolor=blue,%
    urlcolor=blue
} 

\begin{document}
\title{Near isotropic behavior of turbulent thermal convection} 

\author{Dinesh~Nath}
\email{dnath@iitk.ac.in}
\author{Ambrish~Pandey}
\email{ambrishiitk@gmail.com}
\author{Abhishek~Kumar}
\email{abhkr@iitk.ac.in}
\author{Mahendra~K.~Verma}
\email{mkv@iitk.ac.in}
\affiliation{Department of Physics, Indian Institute of Technology Kanpur, Kanpur 208016, India}

\begin{abstract} We investigate the anisotropy in turbulent convection in a 3D box using direct numerical simulation. We compute the anisotropic parameter $A = u_\perp^{2}/(2u_{\parallel}^{2})$, where $u_{\perp}$ and $u_{\parallel}$ are the components of velocity perpendicular and parallel to the buoyancy direction,   the shell and ring spectra, and shell-to-shell energy transfers.  We observe that the flow is nearly isotropic for the Prandtl number $\mathrm{Pr} \approx 1$, but the anisotropy increases with the Prandtl number.  For $\mathrm{Pr}=\infty$, $A \approx 0.3$,   thus anisotropy is not very significant even in extreme cases. We also observe that   $u_{\parallel}$ feeds energy to $u_{\perp}$ via pressure.  The computation of shell-to-shell energy transfers  reveals that the energy transfer in turbulent convection is local and forward, similar to hydrodynamic turbulence.  These results are consistent with the Kolmogorov's spectrum observed by Kumar et al.~[Phys. Rev. E {\bf 90}, 023016 (2014)] for turbulent convection.
\end{abstract}


\maketitle

\section{Introduction} \label{sec:intro}

Kolmogorov's theory of turbulence \cite{Kolmogorov:DANS1941a,Kolmogorov:DANS1941c,Obukhov:DANS1959} describes the flow properties of hydrodynamic turbulence under the assumption of  homogeneity and isotropy of the flow.  This approximation is valid for idealized flows having no  external fields or confining walls.  Most practical flows however involve external fields and walls, hence  they could have significant anisotropy, and their behaviour could differ from  Kolmogorov's theory.  Some of the examples of anisotropic turbulent flows are magnetohydrodynamic (MHD) turbulence \cite{Shebalin:JPP1983}, rotating turbulence~(\cite{Sagaut:book} and references their in), quasi-static liquid-metal flows~\cite{Favier:PF2010a,Favier:PF2010b,Favier:JFM2011,Reddy:PF2014}, turbulent convection~\cite{Rincon:JFM2006}, and rotating convection~\cite{Kunnen:PRL2008}.   In this paper we will study anisotropy in turbulent convection. 

Group theory is used to characterize  properties of system obeying certain symmetries.   Lohse and M\"{u}ller-Groeling~\cite{Lohse:PRE1996}, Kurien and Sreenivasan~\cite{Kurien:PRE2000}, and Biferale and Procaccia~\cite{Biferale:PR2005} expanded the velocity correlation function of isotropic hydrodynamic turbulence using spherical harmonics; under this expansion, the correlation function is spherically symmetric only to the lowest order.  Experiments and numerical simulations~(see review~\cite{Biferale:PR2005} for references) reveal presence of higher-order terms of SO(3) decomposition in the velocity correlation function, thus they demonstrate deviations from the spherical symmetry, a signature of anisotropy.  However, the external gravitational field of the RBC breaks the isotropy of the system and the equation.  Thus, the degree of anisotropy in RBC is expected to be  larger than that in isotropic hydrodynamic turbulence. { Biferale {\em et al.}~\cite{Biferale:PA2004} studied anisotropy in the small-scale turbulence of  convection and random Kolmogorov flow using structure function.  They showed that the anisotropic scaling properties of these two systems are reasonaly similar. }

Rayleigh-B\'{e}nard convection (RBC) is a popular model for studying convective flows~{\color{blue}\cite{Ahlers:PHYSICS2009,Siggia:ARFM1994,Lohse:ARFM2010}}.  RBC is quantified using two parameters: Rayleigh number $\mathrm{Ra}$, which is a measure of the ratio of buoyancy and dissipative force, and Prandtl number $\mathrm{Pr}$, which is the ratio of the kinematic viscosity and thermal diffusivity.  The flow becomes turbulent at large $\mathrm{Ra}$, but it is viscous at very large $\mathrm{Pr}$.  In this paper we  study anisotropy in RBC for a wide range of $\mathrm{Pr}$, from very small value to infinity.  

Researchers have attempted to quantify anisotropy in turbulent flow using several parameters. Shebalin {\em et al.}~\cite{Shebalin:JPP1983} proposed a measure for  a quantity $Q$ in  two-dimensional magnetohydrodynamic turbulence. Some authors (e.g., Sagaut and Cambon~\cite{Sagaut:book} and references therein) used Craya decomposition to quantify toroidal and poloidal components of a vector field.  Teaca {\em et al.}~\cite{Teaca:PRE2009} decomposed the Fourier space into rings and studied energy contents in them, which provides information about the angular dependence of the energy.  They also studied energy exchanges among such rings. For quasi-static MHD, Favier {\em et al.}~\cite{Favier:PF2010a,Favier:PF2010b,Favier:JFM2011} analyzed anisotropy by studying the ratio of the energies of the transverse and parallel components; they quantified the finer details of anisotropy in terms of polar angles  using the toroidal and poloidal  decomposition~\cite{Sagaut:book}. Delache {\em et al.}~\cite{Delache:PF2014} performed similar analysis for rotating turbulence. Reddy and Verma~\cite{Reddy:PF2014} and Reddy {\em et al.}~\cite{Reddy:PP2014}   studied quasi-static MHD using wavenumber rings, and showed how energy is transferred from the parallel component of velocity to the perpendicular component of velocity via pressure.  

Anisotropy in RBC has not been studied in detail. In one of the works, Rincon \cite{Rincon:JFM2006} derived the SO(3) decomposition of structure functions analytically and then computed them numerically.  He showed that the third-order structure function $\langle (\Delta \theta)^2  \Delta {\bf u} \rangle $ appearing in the Yaglom equation exhibits a clear scaling exponent of unity for a small range of scales.  He argued that  buoyancy can deviate the energy spectrum from the Kolmogorov's 5/3 power law.  Kunnen {\em et al.}~\cite{Kunnen:PRL2008}  showed that rotation enhances anisotropy in turbulent convection, with centre plane being {\em rod-like}, while the region near the top plates being nearly isotropic. { Biferale {\em et al.}~\cite{Biferale:PA2004} quantified anisotropy in the small-scale turbulence of  RBC by decomposing the structure function using spherical harmonics.  }

Another class of buoyancy-driven flows is Rayleigh-Taylor instability (RTI)~\cite{Chertkov:PRL2003,Bofetta:JFM1012} that has a very similar instability mechanism as RBC (heavy fluid on top of lighter fluid). Cabot and Zhou~\cite{Cabot:POF2013} and other researchers studied anisotropy  for RTI.  Unstably stratified homogeneous turbulence (USHT) is a class of RTI  in which the integral length of turbulence is much smaller compared to the mixing zone width.  Soulard {\em et al.}~\cite{Soulard:PF2014} and Burlot {\em et al.}~\cite{Burlot:PF2015} studied the phenomenology, spectral modelling, and anisotropy of USHT.  In the Conclusions and Discussions section, we  contrast the observed anisotropy of RBC and USHT.   

In this paper, we perform direct numerical simulation (DNS) of turbulent and laminar RBC flows in a box for different values of $\mathrm{Pr}$ and $\mathrm{Ra}$ and quantify anisotropy in turbulent convection using several measures.  We use anisotropy parameter $A = u_{\perp}^{2}/(2 u_{\parallel}^{2})$, where $u_{\perp}$ and $u_{\parallel}$ are the components of velocity perpendicular and parallel to the buoyancy direction respectively. Next we use ring spectrum to differentiate the energy contents at different polar angles in the Fourier space.  Our simulations reveal nearly isotropic flows, specially for $\mathrm{Pr} \approx 1$.  After this we compute the energy flux and shell-to-shell energy transfers, and show them to be quite close to those observed in hydrodynamic turbulence.  Energetically $u_{\parallel}$ is accelerated by buoyancy.  We show using energy transfer computation that  $u_{\parallel}$ feeds energy to  $u_{\perp}$ via pressure.

The outline of the paper is as follows: In Sec.~\ref{sec:theor-gov-eqs}, we discuss the equations governing the dynamics of Rayleigh-B\'{e}ndard convection. In Sections~\ref{sec:theor-aniso} and ~\ref{sec:theor-Etrans}, we describe respectively the frameworks of anisotropic energy distribution and anisotropic energy transfers used in this paper. We  detail the numerical simulations in Sec.~\ref{sec:sim-method}, and numerical results  of energy spectra and energy transfers in Secs.~\ref{sec:num-E}~and~\ref{sec:num-ETrans} respectively.  Finally, we conclude in Sec.~\ref{sec:conclu}. 

\section{Governing equations} \label{sec:theor-gov-eqs}

The dynamical equations that describe RBC under Boussinesq approximation are~\cite{Chandrasekhar:book} 
\begin{eqnarray}
\frac{\partial \bf u}{\partial t} + (\bf u \cdot \nabla) \bf u & = & - \frac{1}{\rho_0}\nabla \sigma+ \alpha g \theta \hat{z} + \nu\nabla^2 \bf u , \label{eq:u} \\
\frac{\partial \theta}{\partial t} + ({\bf u \cdot \nabla}) \theta & = & \frac{\Delta}{d}u_z+ \kappa \nabla^2 \theta, \label{eq:T} \\
\nabla \cdot \bf u & = & 0 \label{eq:inc}, 
\end{eqnarray}
where $\boldsymbol{\mathrm{u}}$ is the velocity field, $\theta$ and $\sigma$ are the temperature and pressure fluctuations from the conduction state respectively, and $\hat{z}$  is the buoyancy direction. Here $\alpha$ is the thermal expansion coefficient, $g$ is the acceleration due to gravity, and $\rho_0$, $\nu$, $\kappa$ are the fluid's mean density, kinematic viscosity and thermal diffusivity respectively. We  consider fluid between two horizontal plates that are separated by a distance $d$. The temperature difference between the two plates is $\Delta$.  

We nondimensionalize Eqs.~(\ref{eq:u}-\ref{eq:inc}) using $d$ as the length scale, $(\alpha g \Delta d)^{1/2}$ as the velocity scale, and $\Delta$ as the temperature scale, which yields,
\begin{eqnarray}
\frac{\partial{\bf u}}{\partial t}+({\bf u\cdot\nabla){\bf u}} & = & -\nabla\sigma+\theta\hat{z}+\sqrt{\frac{\mathrm{Pr}}{\mathrm{Ra}}}\nabla{}^{2}{\bf u},\label{eq:u_ndim}\\
\frac{\partial\theta}{\partial t}+({\bf u\cdot\nabla)\theta} & = & u{}_{z}+\frac{1}{\sqrt{\mathrm{Ra}\mathrm{Pr}}}\nabla{}^{2}\theta,\label{eq:th_ndim}\\
\nabla\cdot{\bf u} & = & 0.\label{eq:inc_ndim}
\end{eqnarray}
The two nondimensional parameters are the Prandtl number
\begin{equation}
\mathrm{Pr}=\frac{\nu}{\kappa},\label{eq:pr}
\end{equation}
and the Rayleigh number
\begin{equation}
\mathrm{Ra}=\frac{\alpha g\Delta d^{3}}{\nu\kappa}.\label{eq:Ra}
\end{equation}

We solve the aforementioned equations using pseudospectral method in Fourier space.  In our simulations, for the velocity field, we employ the free-slip boundary condition at all the walls.  However, for the temperature field, we employ conducting boundary condition at the top and bottom plates, but insulating boundary condition at the side walls. 

The buoyancy accelerates a fluid parcel along $\hat{z}$, hence it is expected to induce anisotropy in the flow.  Surprisingly, the induced anisotropy  is not very significant.  In the next two sections we will describe tools to characterize the anisotropy in RBC.  

\section{Quantification of anisotropic distribution of energy in RBC} \label{sec:theor-aniso}

In this section, we discuss the anisotropy measures in RBC.  One such measure is the anisotropy parameter $A$ defined as~\cite{Reddy:PF2014}
\begin{equation}
A=\frac{E_{\perp}}{2E_{\parallel}} = \frac{u_\perp^2}{2 u_\parallel^2} =  
\frac{\langle u_x^2 + u_y^2 \rangle}{2 \langle u_z^2\rangle}. \label{eq:Areal}
\end{equation}
Here $\langle . \rangle$ represents averaged quantity per unit volume. For an isotropic flow, $\langle u_x^2 \rangle =  \langle u_y^2 \rangle  = \langle u_z^2 \rangle $, hence $A = 1$.   In RBC,  $\langle u_x^2 \rangle =  \langle u_y^2 \rangle   < \langle u_z^2 \rangle $ because buoyancy preferentially accelerates the flow along $\hat{z}$.  Therefore, $A<1$ for RBC.

\begin{figure}
\begin{centering}
\includegraphics[scale=0.28]{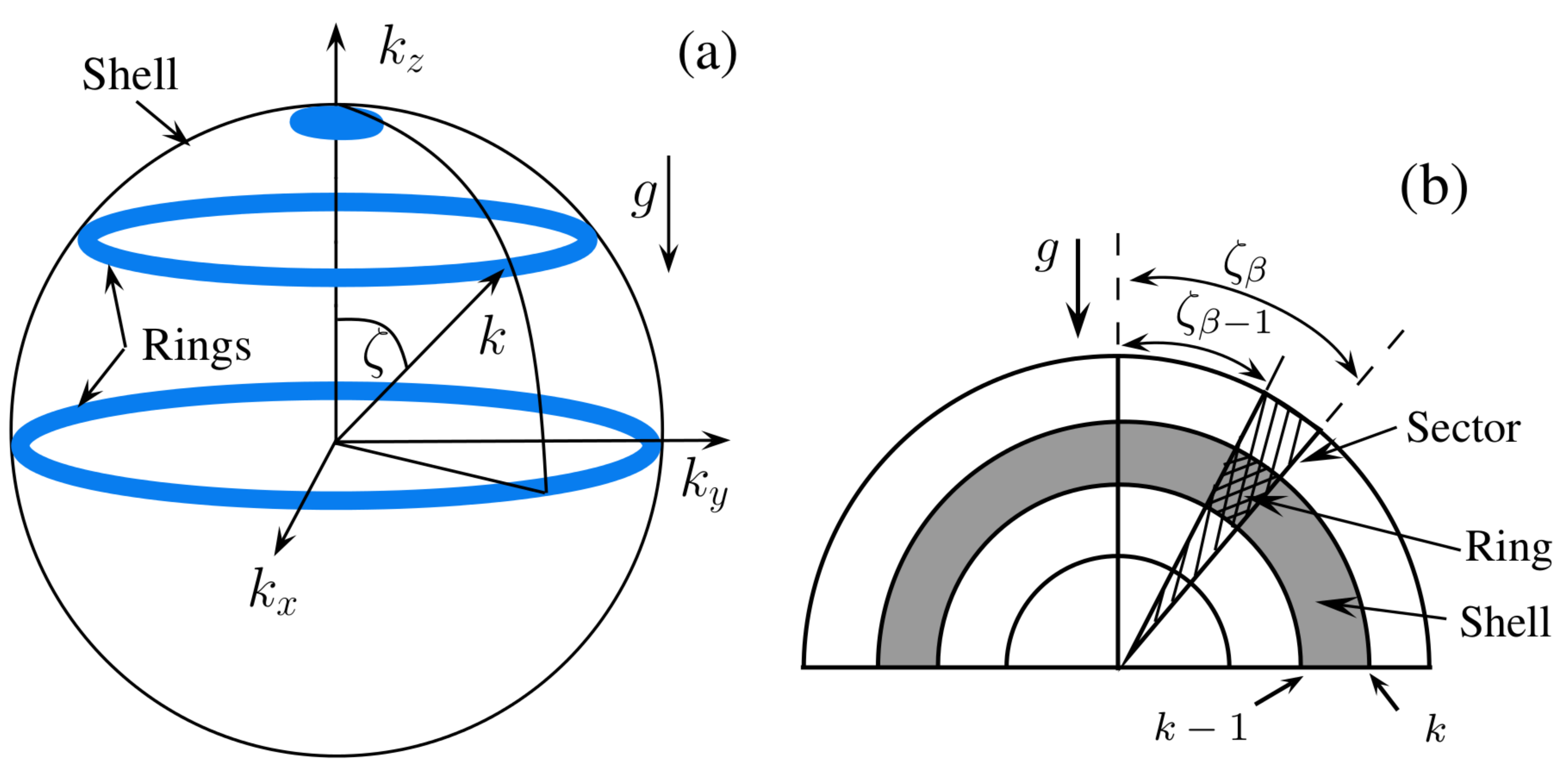}
\end{centering}
\caption{(a) Schematic diagram of ring decomposition in Fourier space. (b) Schematic diagram of a vertical section of the Fourier space exhibiting wavenumber rings, sectors, and shells.}
\label{fig:ring_schem} 
\end{figure}

We quantify the energy at different length scales using the shell spectrum, which is defined as
\begin{equation}
E(k) = \sum_{k-1 < k' \le k} \frac{1}{2}  |{\bf \hat{u}(k')}|^2
\end{equation}
is the sum of the energy of all the Fourier modes [${\bf \hat{u}(k')}$] in a given shell of radius $k$ and unit width.  Thus $E(k)$ masks the  anisotropic features of the flows.  For anisotropic flows, we divide a shell into rings as shown in Fig.~\ref{fig:ring_schem}. Here each shell is divided into rings  that are characterized by two indices---the shell index $k$, and the sector index $\beta$ \cite{Teaca:PRE2009,Reddy:PF2014}.   We name the rings containing $\zeta=0$ and $\zeta=\pi/2$ as the {\em polar} and the {\em equatorial} rings respectively. The gravitational field is aligned along $\zeta=0$.  The energy spectrum of a ring, called the {\em ring spectrum}, is defined as
\begin{equation}
E\left(k,\beta \right)=\frac{1}{C_{\beta}}\sum_{\substack{k-1 < k' \le k;\\
\angle\left(\mathbf{k}'\right)\in\left(\zeta_{\beta-1},\zeta_{\beta}\right]
}
}\frac{1}{2}\left|\mathbf{\hat{u}}\left(\mathbf{k}'\right)\right|^{2},
\label{eq:e_kbeta}
\end{equation}
where $\angle\mathbf{k}'$ is the angle between $\mathbf{k}'$ and the unit vector $\hat{z}$.  The sector $\beta$ contains the modes  between the angles  $\zeta_{\beta-1}$ to $\zeta_{\beta}$ is shown in Fig.~\ref{fig:ring_schem}(b).  For a uniform $\Delta \zeta$, sectors near the equator contain more modes than those near the poles. To compensate the above, we divide the sum $\sum_{k} |{\bf \hat{u}(k')}|^2/2 $ by the factor $C(\beta)$ given by
\begin{equation}
C_{\beta}=\left|\cos\left(\zeta_{\beta-1}\right)-\cos\left(\zeta_{\beta}\right)\right|.
\end{equation}
We obtain further insights into the physics of convection by computing the ring spectra of the perpendicular and parallel components of the velocity:
\begin{eqnarray}
E_\perp(k,\beta) & = & \frac{1}{C_{\beta}}\sum_{\substack{k-1 < k' \le k;\\
\angle\left(\mathbf{k}'\right)\in\left(\zeta_{\beta-1},\zeta_{\beta}\right]
}} \frac{1}{2} \left| \mathbf{\hat{u}}_{\perp} \left(\mathbf{k}'\right) \right|^{2}  \\
E_\parallel(k,\beta) & = & \frac{1}{C_{\beta}}\sum_{\substack{k-1 < k' \le k;\\
\angle\left(\mathbf{k}'\right)\in\left(\zeta_{\beta-1},\zeta_{\beta}\right]
}} \frac{1}{2} \left| \mathbf{\hat{u}}_{\parallel} \left(\mathbf{k}'\right) \right|^{2} 
\end{eqnarray}
where $u_{\parallel}=\mathbf{u}\cdot\hat{z}$ and $\mathbf{u}_{\perp}=\mathbf{u}-u_{\parallel}\hat{z}$.  Clearly, the ring spectrum for the total energy $E(k,\beta) = E_\perp(k,\beta)  + E_\parallel(k,\beta)$.  We can also define energy contents of a sector $\beta$ as
\begin{equation}
E(\beta)= \sum_k E(k,\beta);~~E_{\perp,\parallel}(\beta)= \sum_k E_{\perp,\parallel}(k,\beta).
 \label{eq:e_beta}
\end{equation}
We will compute the above spectra for RBC with various Prandtl and Rayleigh numbers. { We also quantify the anisotropy using Legendre polynomials.  For the same, we expand the ring spectrum of the total energy using Legendre polynomials. }

Earlier Shebalin {\em et al.}~\cite{Shebalin:JPP1983} quantified anisotropy for a quantity $Q$ in  two-dimensional magnetohydrodynamic turbulence using angle $\theta_Q$ that is defined as
\begin{equation}
\theta_Q = \tan^{-1} \frac{\sum_{\bf k} k_z^2 Q({\bf k})}{\sum_{\bf k} k_x^2 Q({\bf k})}.
\end{equation}
In three dimensions, the above formula generalizes to 
\begin{equation}
\theta_Q = \tan^{-1} \frac{\sum_{\bf k} 2k_z^2 Q({\bf k})}{\sum_{\bf k} (k_x^2 +k_y^2)Q({\bf k})}.
\end{equation}
The ring spectrum proposed in this paper provides more detailed information than the angular measure of Shebalin \textit{et al.}~\cite{Shebalin:JPP1983}.

\section{Formalism of energy transfers in RBC} \label{sec:theor-Etrans}
 
The nonlinearity in  fluid flows induces energy transfers among Fourier modes.  In three-dimensional hydrodynamic turbulence, it is known that the energy transfers are forward (from low-wavenumber shells to higher-wavenumber shells) and local (maximal transfer to the neighbouring shell).  In this paper we will investigate whether turbulent convection has a similar behaviour.  We also propose several formulae that capture  anisotropic  energy transfers in RBC.  In the first two subsections, we discuss the energy flux and shell-to-shell energy transfers that describe the isotropic energy exchanges.  In the next two subsections, we describe the ring-to-ring energy transfers and energy exchange between the parallel and perpendicular components of the velocity field.

\subsection{Energy flux} \label{sub:theor-Etrans-flux}

Using Eqs.~(\ref{eq:u},\ref{eq:T}) we derive the equation for the kinetic energy in RBC, which is 
\begin{equation}
\frac{\partial E\left(k\right)}{\partial t}=T\left(k\right)+F\left(k\right)-D\left(k\right),
\label{eq:dEk_dt}
\end{equation}
where $T(k)$ is the energy transfer rate to the shell $k$ due to nonlinear interactions, $F(k)$ is the energy supply rate to the shell due to buoyancy, and $D\left(k\right)$ is the viscous dissipation rate, which are defined as
\begin{eqnarray}
F\left(k\right) & = & \sum_{\left|\mathbf{k'}\right|=k}\Re\langle u_{z}\left(\mathbf{k'}\right)\theta^{*}\left(\mathbf{k'}\right)\rangle,\label{eq:F}\\
D\left(k\right) & = & \sum_{\left|\mathbf{k}'\right|=k}2\sqrt{\frac{\mathrm{Pr}}{\mathrm{Ra}}}k'^{2}E\left(k'\right), \label{eq:D_k}
\end{eqnarray}
where $\Re$ and $\ast$ represent the real part and the complex conjugate of a complex number respectively.   The nonlinear interaction term $T\left(k\right)$ leads to energy transfers among modes leading to a kinetic energy flux $\Pi\left(k_{0 }\right)$, which is defined as
\[
\Pi(k_0)=-\int_{0}^{k_0}T(k)  dk.
\]
Physically, $\Pi(k_0)$ represents the net energy transfers from the modes inside the wavenumber sphere of radius $k_{0}$ to modes outside the sphere~\cite{Pope:book2000}.

The energy flux $\Pi(k_{0})$ is computed using the Fourier space data by employing the mode-to-mode energy transfer~\cite{Dar:PD2001, Verma:PR2004}.  In this framework, we can compute the energy transfers among Fourier modes within an interacting wavenumber triad ${\bf k,p,q}$.  Note that these wavenumbers satisfy a relation ${\bf k = p +q}$.  Here, the rate of energy transfer from the velocity mode  $\mathbf{u(p)}$ to the velocity mode $\mathbf{u(k)}$ with the velocity mode $\mathbf{u(q)}$ acting as a mediator is
\begin{equation}
S\left(\mathbf{k}\left|\mathbf{p}\right|\mathbf{q}\right)=\Im\left\{ \left[\mathbf{k}\cdot\hat{\mathbf{u}}\left(\mathbf{q}\right)\right]\left[\hat{\mathbf{u}}^{\ast}\left(\mathbf{k}\right)\cdot\hat{\mathbf{u}}\left(\mathbf{p}\right)\right]\right\},
\label{eq:mode-to-mode-energy-rate}
\end{equation}
where $\Im$ represents the imaginary part of a complex number. In terms of the above formula, the energy flux is
\begin{equation}
\Pi(k_0) =\sum_{k>k_0} \sum_{p\le k_0} S\left(\mathbf{k}\left|\mathbf{p}\right|\mathbf{q}\right).\label{eq:energy_flux}
\end{equation}

The energy flux provides energy emanating from a wavenumber sphere.  A more detailed picture of the energy transfers is provided by the shell-to-shell energy energy transfers that will be discussed in the next subsection.
 
\subsection{Shell-to-shell energy transfers} \label{sub:shell-to-shell-energy-transfers}

We divide the wavenumber space into a set of wavenumber shells.  The shell-to-shell energy transfer rate from the velocity field of the $m$th shell to the velocity field of the $n$th shell is defined as \cite{Dar:PD2001,Verma:PR2004}
\begin{equation}
T_{n}^{m}=\sum\limits _{\mathbf{k}\in n}\sum\limits _{\mathbf{p}\in m}S\left(\mathbf{k}\left|\mathbf{p}\right|\mathbf{q}\right).\label{eq:shell-to-shell-energy-rate}
\end{equation}
In Kolmogorov's theory of hydrodynamic turbulence, the maximum shell-to-shell energy transfer occurs from shell $m$ to shell $(m+1)$, hence the energy transfer in hydrodynamic turbulence is {\em local} and {\em forward}.  We will explore using numerical simulations whether  the energy transfer in RBC is local and forward.

Unfortunately both the aforementioned measures of energy transfers do not capture the anisotropic effects since the energy flux and shell-to-shell energy transfers provide averaged values over the polar angle $\zeta$.  In the next two subsections we quantify anisotropic energy transfers using ring-to-ring energy transfers and the energy exchange  between the perpendicular and parallel components of the velocity field.

\subsection{Ring-to-ring energy transfers}

As discussed in Sec.~\ref{sec:theor-aniso}, the Fourier space is divided into rings.  The nonlinear interactions  among the Fourier modes induce energy transfers among the rings.  The ring-to-ring energy transfer from ring $(m,\alpha)$ to  $(n,\beta)$ is
\begin{equation}
T_{(n,\alpha)}^{(n,\beta)}=\sum\limits _{\mathbf{k}\in (n,\beta)}\sum\limits _{\mathbf{p}\in (m,\alpha)}S\left(\mathbf{k}\left|\mathbf{p}\right|\mathbf{q}\right).\label{eq:ring-to-ring-energy-rate}
\end{equation}
These calculations for all $m$ and $n$'s are very detailed and time-consuming since they involve a large number of rings.

\subsection{Energy exchange between perpendicular and parallel components of the velocity field}
\label{sub:theor-Etrans-para-perp}

Since buoyancy accelerates a fluid parcel along the vertical direction ($\hat{z}$), the energy  of the parallel component of velocity ($E_\parallel = \langle u_\parallel^2/2 \rangle$ or $\langle u_z^2/2 \rangle$) exceeds the energy  of any of the two horizontal components ($\langle u_x^2/2 \rangle$  or $\langle u_y^2/2 \rangle$).  The horizontal components of the velocity however is forced by the pressure gradient and the viscous force, the latter being a dissipative.      We show below how $u_\parallel$ feeds energy to  ${\bf u}_\perp$. 
   
Following the procedure used by Reddy \textit{et al.}~\cite{Reddy:PP2014}, the energy equations for the perpendicular and parallel components of the velocity field are as follows:
\begin{eqnarray}
\frac{\partial E_{\perp}\left(\mathbf{k}\right)}{\partial t} & = & \sum\limits _{\mathbf{k}=\mathbf{p}+\mathbf{q}}S_{\perp}\left(\mathbf{k}\left|\mathbf{p}\right|\mathbf{q}\right)+P_{\perp}\left(\mathbf{k}\right) -2\sqrt{\frac{\mathrm{Pr}}{\mathrm{Ra}}}\sum\limits _{\left|\mathbf{k}\right|=k}k^{2}E_{\perp}\left(\mathbf{k}\right),\label{eq:Eperp_t_eqn}\\
\frac{\partial E_{\parallel}\left(\mathbf{k}\right)}{\partial t} & = & \sum\limits _{\mathbf{k}=\mathbf{p}+\mathbf{q}}S_{\parallel}\left(\mathbf{k}\left|\mathbf{p}\right|\mathbf{q}\right)+P_{\parallel}\left(\mathbf{k}\right) -2\sqrt{\frac{\mathrm{Pr}}{\mathrm{Ra}}}\sum\limits _{\left|\mathbf{k}\right|=k}k^{2}E_{\parallel}\left(\mathbf{k}\right) +\sum\limits _{\left|\mathbf{k}\right|=k}\Re\left(\langle u_{\parallel}\left(\mathbf{k}\right)\theta^{\ast}\left(\mathbf{k}\right)\rangle\right),\label{eq:Epara_t_eqn}
\end{eqnarray}
where
\begin{eqnarray}
S_{\perp}\left(\mathbf{k}\left|\mathbf{p}\right|\mathbf{q}\right) & = & \Im\left\{ \left[\mathbf{k}\cdot\hat{\mathbf{u}}\left(\mathbf{q}\right)\right]\left[\hat{\mathbf{u}}_{\perp}^{\ast}\left(\mathbf{k}\right)\cdot\hat{\mathbf{u}}_{\perp}\left(\mathbf{p}\right)\right]\right\} , \label{eq:S_perp}\\
S_{\parallel}\left(\mathbf{k}\left|\mathbf{p}\right|\mathbf{q}\right) & = & \Im\left\{ \left[\mathbf{k}\cdot\hat{\mathbf{u}}\left(\mathbf{q}\right)\right]\left[\hat{u}_{\parallel}^{\ast}\left(\mathbf{k}\right)\hat{u}_{\parallel}\left(\mathbf{p}\right)\right]\right\},  \label{eq:S_para}\\
P_{\perp}\left(\mathbf{k}\right) & = & \Im\left\{ \left[\mathbf{k}_{\perp}\cdot\hat{\mathbf{u}}_{\perp}^{\ast}\left(\mathbf{k}\right)\right]\hat{\sigma}\left(\mathbf{k}\right)\right\}, \label{eq:pperp}\\
P_{\parallel}\left(\mathbf{k}\right) & = & \Im\left\{ \left[k_{\parallel}\hat{u}_{\parallel}^{\ast}\left(\mathbf{k}\right)\right]\hat{\sigma}\left(\mathbf{k}\right)\right\}. \label{eq:ppara}
\end{eqnarray}
Reddy \textit{et al.}~\cite{Reddy:PP2014} showed that  $S_{\perp}\left(\mathbf{k}\left|\mathbf{p}\right|\mathbf{q}\right)$ as well as $S_{\parallel}\left(\mathbf{k}\left|\mathbf{p}\right|\mathbf{q}\right)$ are invidually conserved in a triadic interaction.  Following the line of arguments of Dar \textit{et al.}~\cite{Dar:PD2001}, Verma \cite{Verma:PR2004}, and Reddy \textit{et al.}~\cite{Reddy:PP2014}, we can define energy fluxes of the perpendicular and parallel components of the velocity field as 
\begin{eqnarray}
\Pi_{\perp}\left(k'\right) & = &\sum\limits _{\left|\mathbf{k}\right|\geq k'}\sum\limits _{\left|\mathbf{p}\right|<k'}S_{\perp}\left(\mathbf{k}\left|\mathbf{p}\right|\mathbf{q}\right) \\
\Pi_{\parallel}\left(k'\right) & = &\sum\limits _{|\mathbf{k}|\geq k'}\sum\limits _{|\mathbf{p}|<k'}S_{\parallel}\left(\mathbf{k}\left|\mathbf{p}\right|\mathbf{q}\right).
\end{eqnarray}
It is important to note that there is no direct energy transfer from the parallel component to the perpendicular component, i.e., the nonlinear transfer does not have terms of the type $[\hat{\mathbf{u}}_{\perp}^{\ast}\left(\mathbf{k}\right)\cdot\hat{\mathbf{u}}_{\parallel}\left(\mathbf{p}\right) ] $ (which would anyway vanish since they are perpendicular to each other).

The energy equations [Eqs.~(\ref{eq:Eperp_t_eqn},\ref{eq:Epara_t_eqn})] reveal that $E_{\perp}$ receives  energy from $E_\parallel$  by an amount $P_{\perp}\left(\mathbf{k}\right)$.   Interestingly, the incompressibility condition [$\mathbf{k}\cdot\hat{\mathbf{u}}\left(\mathbf{k}\right)=0$] yields
\begin{equation}
P_{\perp}\left(\mathbf{k}\right)=-P_{\parallel}\left(\mathbf{k}\right).
\label{eq:P}
\end{equation}
That is, the energy gained by the  perpendicular component $\hat{u}_{\perp}^{\ast}\left(\mathbf{k}\right)$ via pressure is equal and opposite to the energy lost by the  parallel counterpart. The pressure aids the energy exchange between the perpendicular and parallel components of velocity field.   The total flux $\Pi$ is the sum of $\Pi_{\perp}$ and $\Pi_{\parallel}$.

We will study the above transfers for RBC in Sec.~\ref{sec:num-ETrans} using numerical data.  In the next section we will describe simulation details.

\section{Simulation details} \label{sec:sim-method}

We perform direct numerical simulation of RBC in a cube of unit size (aspect ratio one) using a pseudo-spectral solver {\sc Tarang} \cite{Verma:Pramana2013}. We employ a $512^3$ grid for our simulations.  The nondimensional equations [Eqs.~(\ref{eq:u_ndim}--\ref{eq:inc_ndim})] are evolved in time by the fourth order Runge-Kutta method. The CFL (Courant-Friedrichs-Lewy) condition is used for calculating the time step $\Delta t$ (a typical value is $0.01$), and the $3/2$ rule for the dealiasing. 

\begin{table}
\caption{RBC simulation parameters with grid resolution of $512^3$: Prandtl number $\mathrm{Pr}$, Rayleigh number $\mathrm{Ra}$, Reynolds number $\mathrm{Re}$, $k_{max}\eta$,  $E_{\perp}/2$, $E_{\parallel}$,  anisotropic parameter $A = E_{\perp}/(2E_{\parallel})$, integral length scale $l$, vertical integral length scale $l_z$, and the total viscous dissipation rate $D$.}
\begin{ruledtabular}
\begin{tabular}{cccccccccc}
$\mathrm{Pr}$ & $\mathrm{Ra}$ & $\mathrm{Re}$ & $k_\mathrm{max}\eta$ & $0.5E_{\perp}$ & $E_{\parallel}$ & $A$ & $l$ & $l_{z}$ & $D$\tabularnewline
\hline 
$0.02$ & $2\times10^{6}$ & $7.05\times10^{3}$ & $3.4$ & $0.118$ & $0.187$ & $0.63$ & $0.478$  & $0.553$ & $1.02\times10^{-4}$\tabularnewline
$1$ & $10^{8}$ & $3.11\times10^{3}$ & $5.9$ & $0.013$ & $0.018$ & $0.73$ & $0.468$  & $0.547$ & $1.02\times10^{-4}$\tabularnewline
$6.8$ & $10^{8}$ & $9.08\times10^{2}$ & $3.2$ & $0.010$ & $0.017$ & $0.59$ & $0.484$ & $0.591$ & $2.65\times10^{-4}$\tabularnewline
$100$ & $10^{8}$ & $1.25\times10^{2}$ & $1.6$ & $0.002$ & $0.004$ & $0.49$ & $0.531$  & $0.635$ & $1.02\times10^{-3}$\tabularnewline
$\infty$ & $2\times10^{8}$ & $0$ & $4.2$ & $0.221$ & $0.725$ & $0.30$ & $0.449$ & $0.716$ & $7.21\times10^{-5}$\tabularnewline
\end{tabular}
\end{ruledtabular}
\label{tab:sim-para}
\end{table}

\begin{figure}
\includegraphics[scale=0.27]{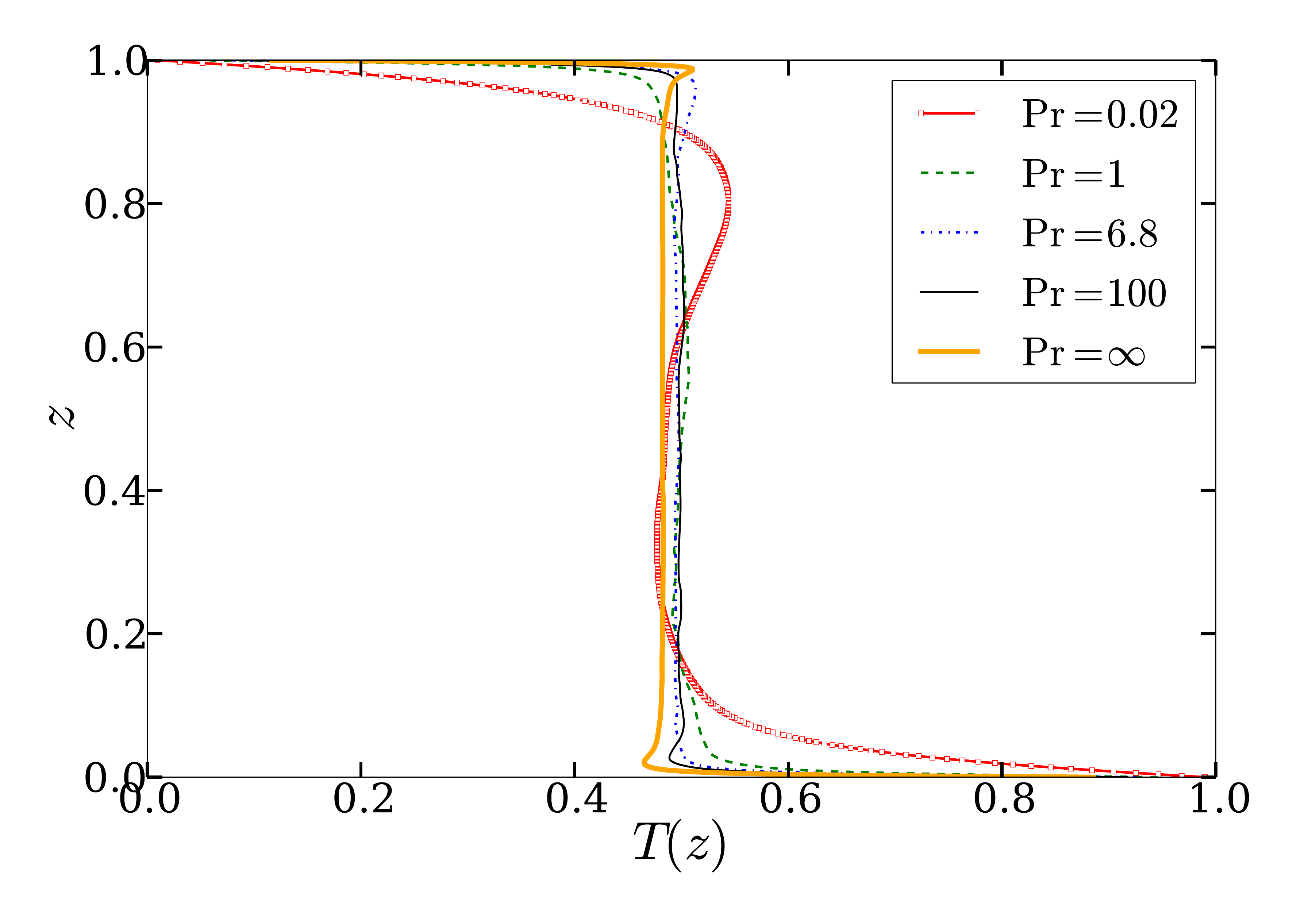}
\caption{{ Plot of the instantaneous temperature profile $T(z) =  \int dx dy T(x,y,z) $ vs. $z$. that exhibits $T(z) \approx 1/2$ in the bulk, and sharp variations near the walls.}}
\label{fig:temp_profile}
\end{figure}

The boundary conditions employed in this work are as follows: free-slip boundary condition for the velocity  at all the walls, conducting boundary condition for the temperature at the top and bottom walls, and insulating boundary condition for the temperature at the vertical walls.  We remark that the energy spectra and energy transfer studies are best performed in the Fourier space.  The free-slip boundary condition facilitates transformation to the Fourier space quite naturally, in contrast to the no-slip boundary condition for which one needs to invoke Chebyshev polynomials.  This is one of the reasons for choosing the free-slip boundary condition for our simulation.  Note that the energy spectrum and energy transfers in the bulk of the flow are expected to be similar for the free-slip and no-slip boundary conditions. { In Fig.~\ref{fig:temp_profile}, we show the (horizontal) planar-averaged temperature $T(z)$ as a function of $z$, where $T(z) = \int dx dy T(x,y,z)$. The figure shows that $T(z)$ has a sharp gradient near the walls, and an approximate constant value ($\approx 1/2$) in the bulk.}

We initiate an RBC simulation  on a smaller grid  and continue it until the steady state is reached. The final state of this simulation is used as an initial condition for a simulation on a larger grid for a set of larger parameters ($\mathrm{Pr}$ and $\mathrm{Ra}$).   The final run is performed on $512^{3}$ grid size. The various values of $\mathrm{Pr}$ and $\mathrm{Ra}$ considered in this work are listed in Table~\ref{tab:sim-para}. In each simulation, the quantity $k_\mathrm{max}\eta$  remains greater than unity  implying that the simulations are well resolved. Here $k_\mathrm{max}$ is the maximum wavenumber, { and $\eta$ represents the Batchelor length ($\eta_\theta$), except for $\mathrm{Pr} = 0.02$, for which $\eta$ represent the Kolmogorov length ($\eta_u$).}  Since the box size $d=1$, the wavenumber ${\bf k} = \pi(n_x, n_y, n_z)$, where $n_x, n_y, n_z$ are integers. Therefore, $\Delta k_x = \Delta k_y = \Delta k_z = \pi$,  the smallest wavevectors in system  are ${\bf k} = \pi(\hat{x}+\hat{z})$ and ${\bf k} = \pi(\hat{y}+\hat{z})$, and hence $k_\mathrm{min} = \pi\sqrt{2}$.  

We compute the integral length $l$, and vertical integral length $l_z$ using the formulae:
\begin{eqnarray}
l & = &  \frac{\sum_k E(k)(\pi/k) }{ \sum_k E(k) }  \\
l_z & = &  \frac{ \sum_k \sum_\beta E(k,\beta)(\pi/k_z) }{ \sum_k \sum_\beta  E(k,\beta) }  
\end{eqnarray}
where $k_z = k \cos \zeta$ with $\zeta$ as the mean angle in the sector $\beta$.  The computed values of $l$ and $l_z$, listed in Table~\ref{tab:sim-para}, are comparable to the box size.  Note that $l_z > l$ with the maximum $l_z$ occurring  for $\mathrm{Pr}=\infty$, and minimum for $\mathrm{Pr}=1$, which is connected to the elongation of the plumes for larger $\mathrm{Pr}$.   Thus, the least anisotropic flow occurs for $\mathrm{Pr}=1$, and most anisotropic one for $\mathrm{Pr}=\infty$.  This feature will be discussed  in the following sections.

\begin{figure}[tp]
\begin{centering}
\includegraphics[scale=0.52]{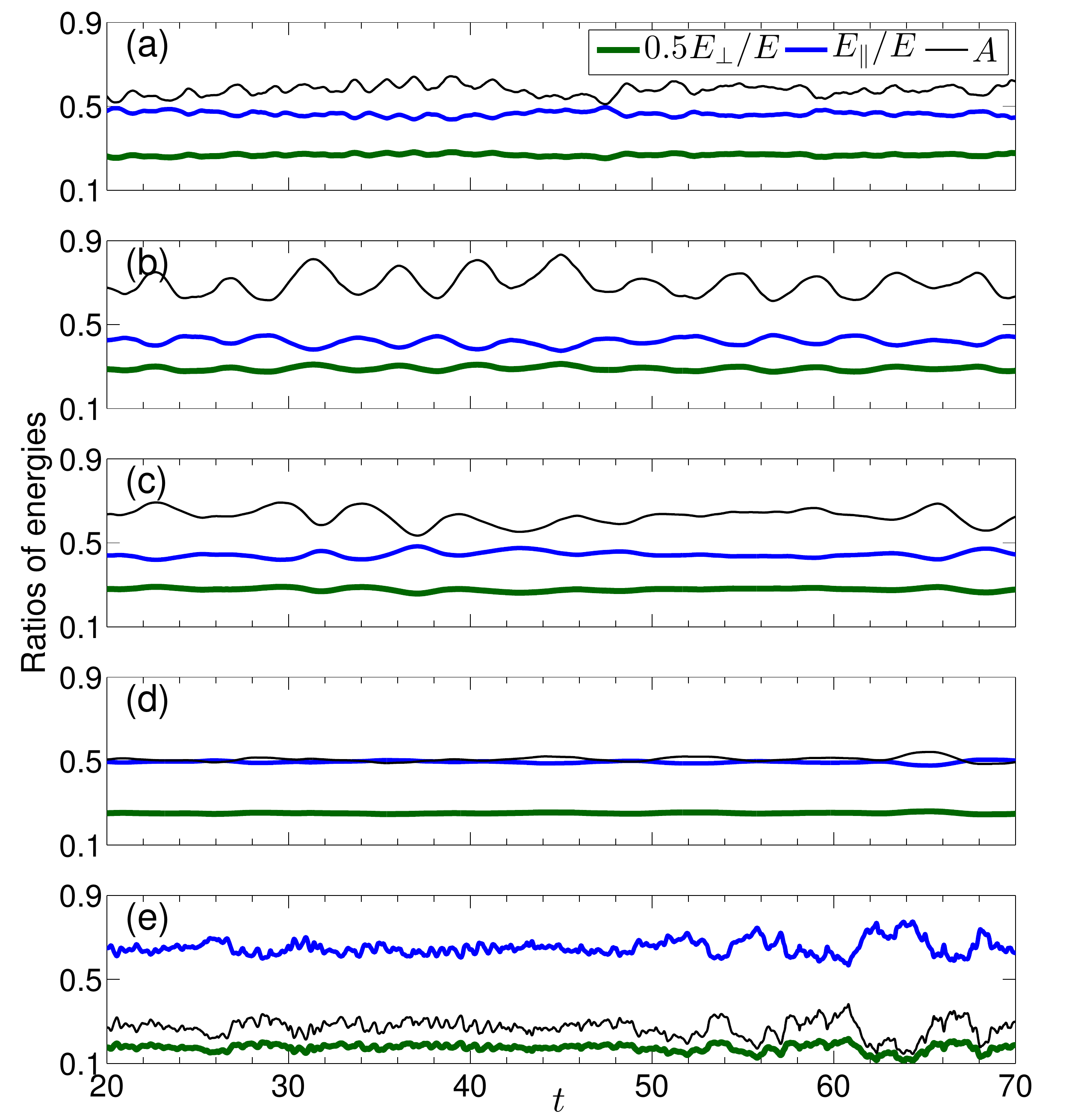}
\end{centering}
\caption{Time series of $E_{\perp}/(2E)$, $E_{\parallel}/E$, and $A = E_{\perp}/(2E_{\parallel})$ for (a)
$\mathrm{Pr=0.02}$, $\mathrm{Ra=2\times10^{6}}$, (b) $\mathrm{Pr=1}$, $\mathrm{Ra=10^{8}}$, (c) $\mathrm{Pr=6.8}$, $\mathrm{Ra=10^{8}}$, (d) $\mathrm{Pr=100}$, $\mathrm{Ra=10^{8}}$, and (e) $\mathrm{Pr=\infty}$, $\mathrm{Ra=2\times10^{8}}$.  }
\label{fig:timeEvolution} 
\end{figure}

\section{Numerical results on energy distribution in RBC} \label{sec:num-E} 

In this section we describe the energy distribution in turbulent RBC for various Prandtl numbers.  In Fig.~\ref{fig:timeEvolution} we exhibit the time series of  the steady-state values of $E_\perp/(2E)$, $ E_\parallel/E$, and the anisotropy parameter $A = E_\perp/(2E_\parallel)$ for $\mathrm{Pr} = 0.02, 1, 6.8, 100$, and $ \infty$.  The average values of $A$ for these parameters are $0.63, 0.73, 0.59, 0.49$ and $0.30$ respectively (listed in Table~\ref{tab:sim-para}), consistent with the argument that buoyancy yields $u_\parallel^2 > (u_\perp^2/2)$.  These results demonstrate that the flow is least anisotropic (maximum $A$) for $\mathrm{Pr} =1$, and most anisotropic (minimum $A$) for $\mathrm{Pr} =\infty$, consistent with our computations of integral length scales.   Note however that in the extreme case, $\mathrm{Pr} =\infty$, $E_y$  and $E_x$ are approximately 30\% of $E_z$.   Hence,  the flow  is not far from isotropy even for extreme cases (large and infinite Prandtl numbers).   Table~\ref{tab:sim-para} also lists the total viscous dissipation rate, which is quite small for all our runs.

\begin{figure}
\includegraphics[scale=0.61]{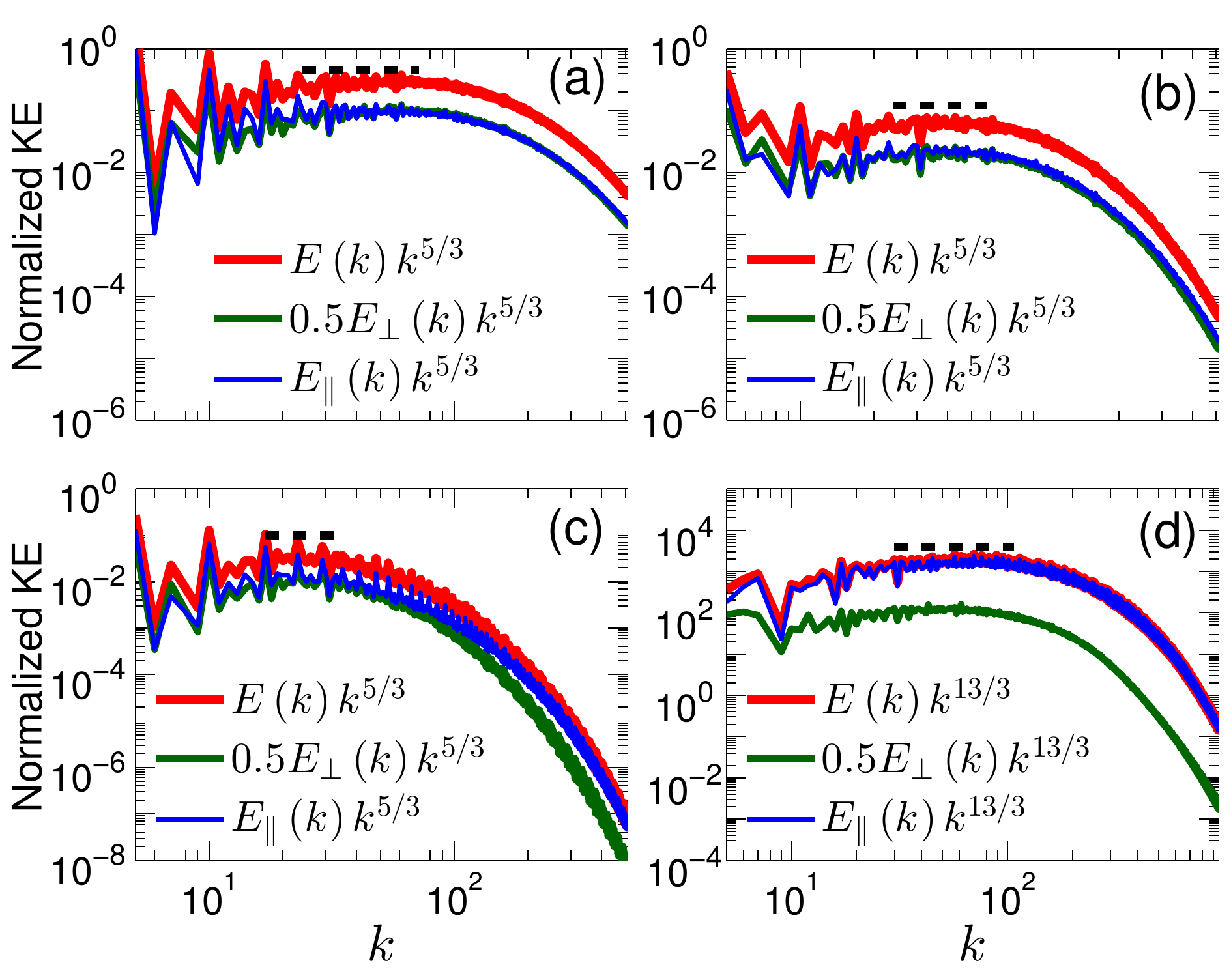}
\caption{Plots of normalized kinetic energy spectrum $E(k) k^{\alpha}$ vs.~$k$ for (a) $\mathrm{Pr}=0.02$, $\mathrm{Ra}=2\times10^{6}$, (b) $\mathrm{Pr}=1$, $\mathrm{Ra}=10^{8}$, (c) $\mathrm{Pr}=6.8$, $\mathrm{Ra}=10^{8}$, and (d) $\mathrm{Pr}=\infty$, $\mathrm{Ra}=2\times10^{8}$. Black dashed lines depict the power-law regimes.}
\label{fig:Ek_vs_k} 
\end{figure}

Next, we compute the energy spectra $E(k)$, $E_\perp(k)$, and $E_\parallel(k)$ for $\mathrm{Pr} = 0.02, 1, 6.8, \infty$.   We observe that $E_\parallel(k) > E_\perp(k)/2$ for all the cases, with the divergence between  $E_\parallel(k)$ and $ E_\perp(k)/2$ being maximum for $\mathrm{Pr}=\infty$ (see Fig.~\ref{fig:Ek_vs_k}).  For a small inertial range, we observe $E(k) \sim k^{-5/3}$  for $\mathrm{Pr} = 0.02, 1, 6.8$, and $E(k) \sim k^{-13/3}$ for $\mathrm{Pr} = \infty$, consistent with the earlier results of Kumar \textit{et al.}~\cite{Kumar:PRE2014} and Pandey \textit{et al.}~\cite{Pandey:PRE2014} respectively.   The power-law regime is rather small due to relatively smaller grid resolutions in our simulations.

\begin{figure}
\includegraphics[scale=0.25]{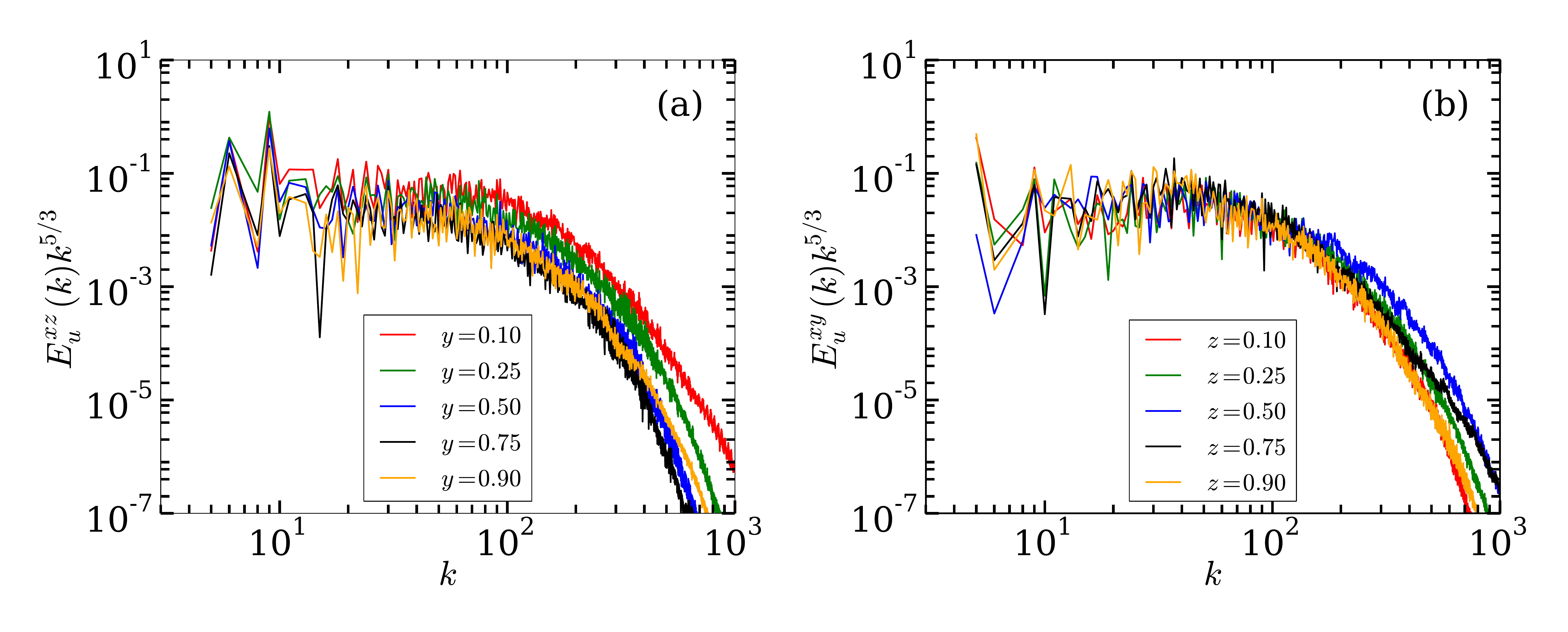}
\caption{{ For $\mathrm{Pr} = 1, \mathrm{Ra} = 10^8$, plots of two-dimensional normalized kinetic spectra: (a) $E^{xz}_u(k)k^{5/3}$ for vertical sections at $y=0.10, 0.25, 0.5, 0.75, 0.90$,  and (b) $E^{xy}_u(k)k^{5/3}$ for the horizontal sections at $z=0.10, 0.25, 0.5, 0.75, 0.90$. The near overlapping plots demonstrate  near homogeneity and isotropy of the flow. }}
\label{fig:2d_spectrum}
\end{figure}

 { To explore whether the flow is homogeneous and isotropic, we take various horizontal and vertical sections of the flow at $y=0.10, 0.25, 0.5, 0.75, 0.90$ and $z=0.10, 0.25, 0.5, 0.75, 0.90$. We compute the normalized two-dimensional energy spectra for these sections for $k$ in the inertial range.  In Fig.~\ref{fig:2d_spectrum}(a,b) we present these spectra for $\mathrm{Pr} = 1, \mathrm{Ra} = 10^8$.   The plots nearly overlap with each other, thus demonstrating near homogeneity and isotropy of the flow.}

\begin{figure}
\includegraphics[scale=0.8]{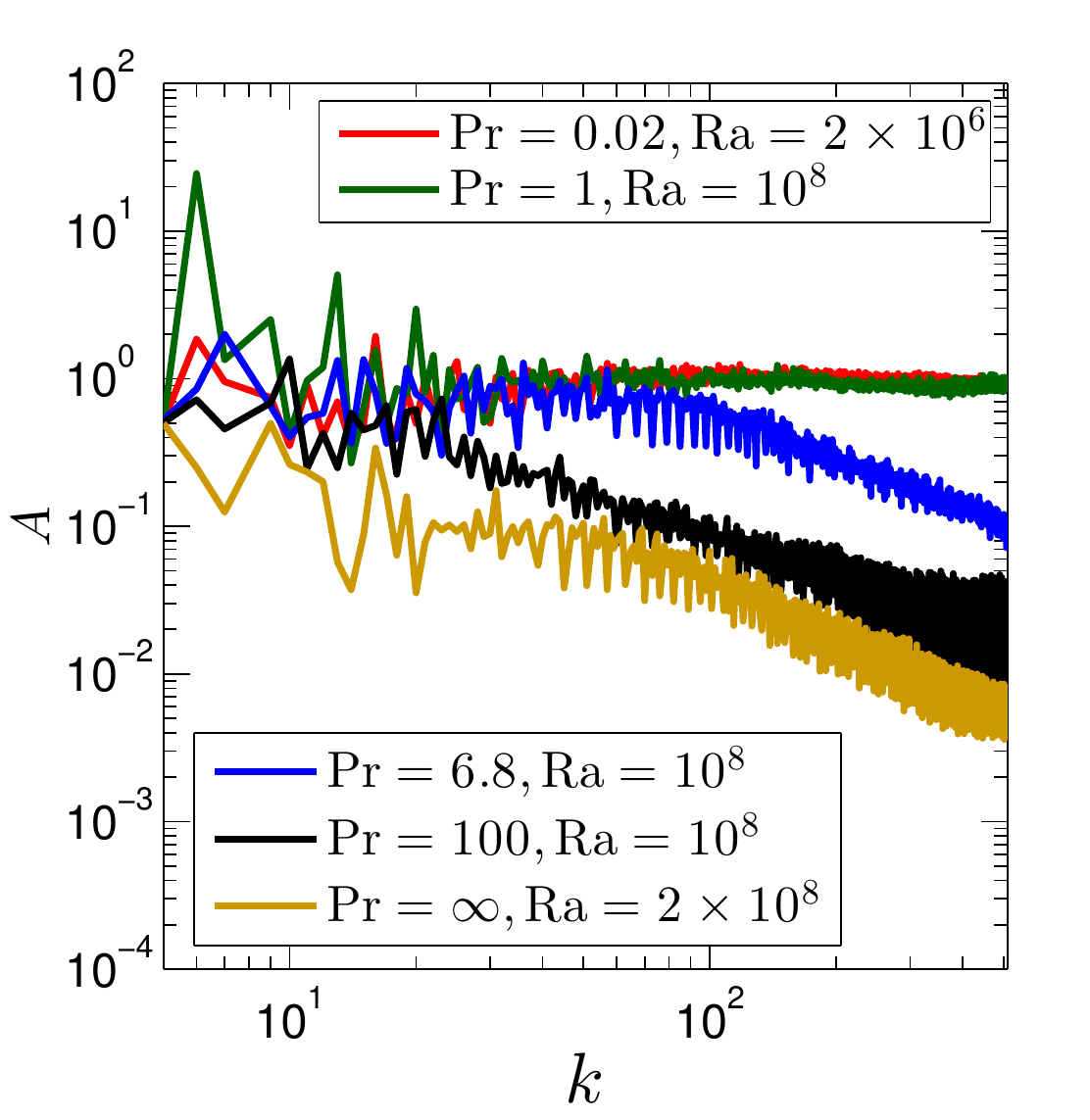}
\caption{Plot of $A(k) = E_\perp(k)/(2E_\parallel(k))$ vs.~$k$ for  $\mathrm{Pr} = 0.02, 1, 6.8, 100$, and $ \infty$. For large $k$, $A(k) \ll 1$ for $\mathrm{Pr} = 100$ and $ \infty$. }
\label{fig:A(k)} 
\end{figure}

In Fig.~\ref{fig:A(k)} we plot the anisotropic parameter $A(k) = E_\perp(k)/(2E_\parallel(k))$ vs.~$k$ that quantifies the scale-by-scale anisotropy.  We observe that for large $k$, $A(k) \approx 1$ for $\mathrm{Pr} = 0.02, 1$.  For large and infinite $\mathrm{Pr}$, $A(k) \ll 1$ with $A(k)$ decreasing with $k$ after $k>50$ or so.   Thus, the flows with  large and infinite $\mathrm{Pr}$ are strongly anisotropic  at small scales ($k \gg 1$), a feature related to the thin plumes.  We will revisit this aspect while discussing the ring spectrum.

\begin{figure}
\includegraphics[scale=0.75]{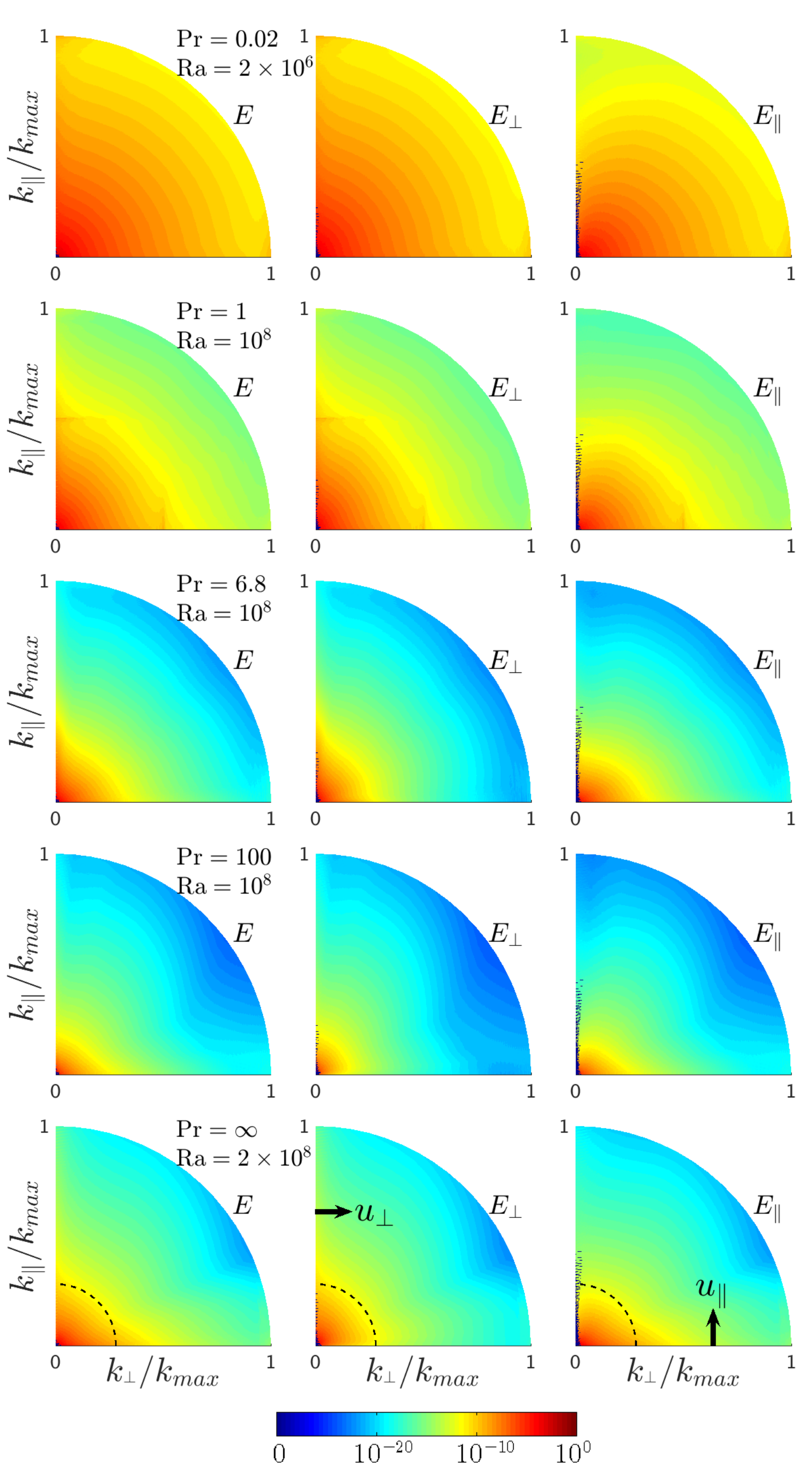}
\caption{Ring spectra  $E(k,\beta)$, $E_{\perp}(k,\beta)$ and $E_{\parallel}(k,\beta)$ (from left to right) depicted for various $\mathrm{Pr}$ and $\mathrm{Ra}$ (from top to bottom).  The plots show nearly isotropic distribution of energy. }
\label{fig:Ek-density} 
\end{figure}

Now we compute the ring spectra $E(k,\beta)$, defined in Sec.~\ref{sec:theor-aniso}, for various Prandtl numbers.  For this we have divided the Fourier space into shells of unit radius from $k=0$ to $k_\mathrm{max}=\pi N$ (grid size = $N^3$), and 20 sectors with  $\Delta \zeta$ of $4.5$ degrees each.  In Fig.~\ref{fig:Ek-density} we exhibit the ring spectra for $\mathrm{Pr} = 0.02, 1, 6.8, 100$, and $\infty$, with $z$ axis along the vertical.  The plots indicate that the energy distribution is nearly isotropic.   However, some interesting deviations from isotropy  are as follows:

\begin{enumerate}

\item For the Prandtl number near unity, the deviations from isotropy is strongest near the diagonal region.  This is primarily because of the dominance of   large-scale rolls for which $(k_x=1, k_y = 0, k_z=1)$ and $(k_x=0, k_y = 1, k_z=1)$ are the most important Fourier modes.  The large-scale secondary modes generated by nonlinearity are expected to have $k_x \approx k_z \gg k_y$ or $k_y \approx k_z \gg k_x$.  These modes have similar structures, and they contribute more to the rings near $\zeta \approx \pi/4$.  

\item On comparison of anisotropy for different Prandtl numbers, we observe that  the anisotropy is maximal for $\mathrm{Pr} =100$ and $\infty$, specially near the $\zeta=0$ and $\zeta=\pi/2$.   We can attribute this anisotropy to the strong plume structures observed in RBC with large Pr.  The incompressibility condition
\begin{equation}
{\bf k \cdot u(k)} = k_\parallel u_\parallel + {\bf k_\perp \cdot u_\perp(k)} = 0
\end{equation}
implies that $u_\parallel \gg u_\perp$ when $k_\parallel \ll k_\perp$.  Consequently, the ascending and descending plumes have uniform structures along $z$ but strong variability along the horizontal (due to their thin structures).  The other regime $k_\perp \ll k_\parallel$ has $u_\perp \gg u_\parallel$  that corresponds to the horizontal flow near the top and bottom plates.  Thus, the incompressibility plays a major role in the determination of anisotropic structures  in RBC.

\end{enumerate}

\begin{figure}
\includegraphics[scale=0.60]{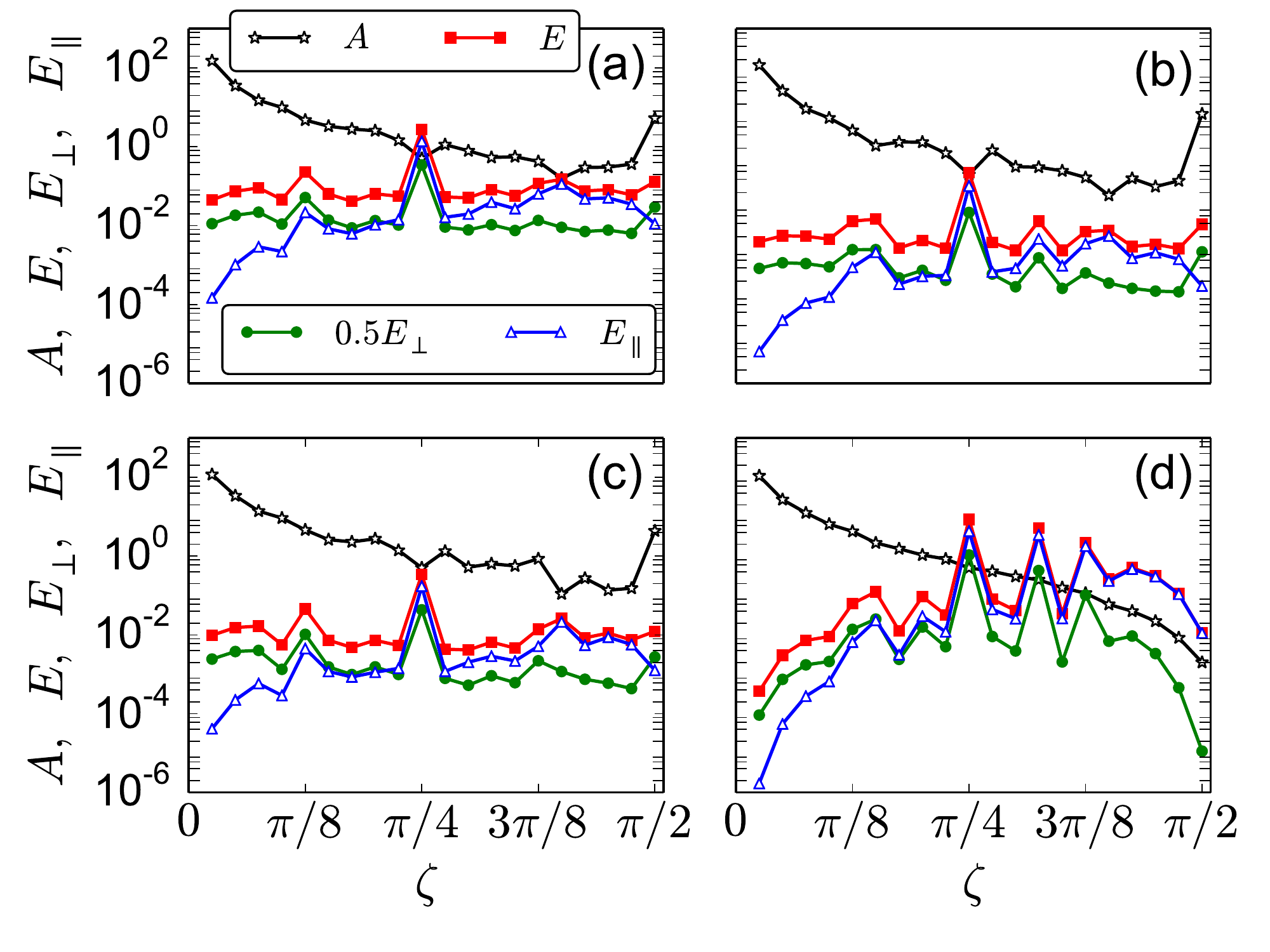}
\caption{Plots of sector energy $E(\beta)$,  $E_\perp(\beta)$, and $E_\parallel(\beta)$  [see  Eq.~(\ref{eq:e_beta})] as a function angle $\zeta$ for (a) $\mathrm{Pr}=0.02$, $\mathrm{Ra}=2\times10^{6}$,
(b) $\mathrm{Pr}=1$, $\mathrm{Ra}=10^{8}$, (c) $\mathrm{Pr}=6.8$, $\mathrm{Ra}=10^{8}$, and (d) $\mathrm{Pr}=\infty$, $\mathrm{Ra}=2\times10^{8}$. The legends for (b--d) are same as in (a).}
\label{fig:E_zeta}
\end{figure}

To highlight the angular dependence of the ring spectrum,  in Fig.~\ref{fig:E_zeta} we plot the energy in the sector  $\beta$, $E(\beta) $, $E_\perp(\beta)$, and $E_\parallel(\beta)$ as a function of $\zeta$ [refer to Eq.~(\ref{eq:e_beta})]. The plots indicate a peak in these spectra near $\zeta = \pi/4$, consistent with the fact that the maximum of the energy spectrum occurs near the diagonal region. We also observe strong $E_\perp(k,\beta)$ and $E_\parallel(k,\beta)$ for $\zeta \approx 0$ and $\zeta \approx \pi/2$ respectively, consistent with item (2) discussed above.  

{
We can also quantify anisotropy by expanding $E(k,\zeta)$ using Legendre polynomials as
\begin{equation}
E(k,\zeta) = \sum_l a_l(k) P_l(\cos \zeta).
\label{eq:Legendre}
\end{equation}
For isotropic distribution, we expect $a_0 \ne 0$, while the other $a_l$ should be relatively small. However for the strongly anisotropic flows, the amplitudes of $a_l$ for higher $l$'s dominate $a_0$.  See for example, quasi two-dimensional flow of quasi-static magnetohydrodynamic flow reported by Reddy and Verma~\cite{Reddy:PF2014}.   We compute $a_l$ by inverting Eq.~(\ref{eq:Legendre}) for a particular wavenumber in the inertial range.  We find that the $a_l$ for odd $l$'s nearly vanish due to $\zeta \rightarrow \pi - \zeta$ symmetry.  In Fig.~\ref{fig:legendre} we plot $|a_{l}|$ vs.~$l$; these coefficients have similar amplitudes indicating deviations from isotropy.  Stronger values of $E(k,\zeta)$ near the equator and polar regions are due to the nonzero values of $a_{l}$ with $l > 0$. Note however that  $a_l$ for large $l$ is not very large compared to $a_0$~\cite{Reddy:PF2014}, hence the flow is not strongly anisotropic.  Our description of anisotropy in the ring spectrum using Legendre polynomial is complimentary to that of Biferale {\em et al.}~\cite{Biferale:PA2004,Biferale:PR2005} who quantify anisotropy using structure function.
\begin{figure}
\includegraphics[scale=0.3]{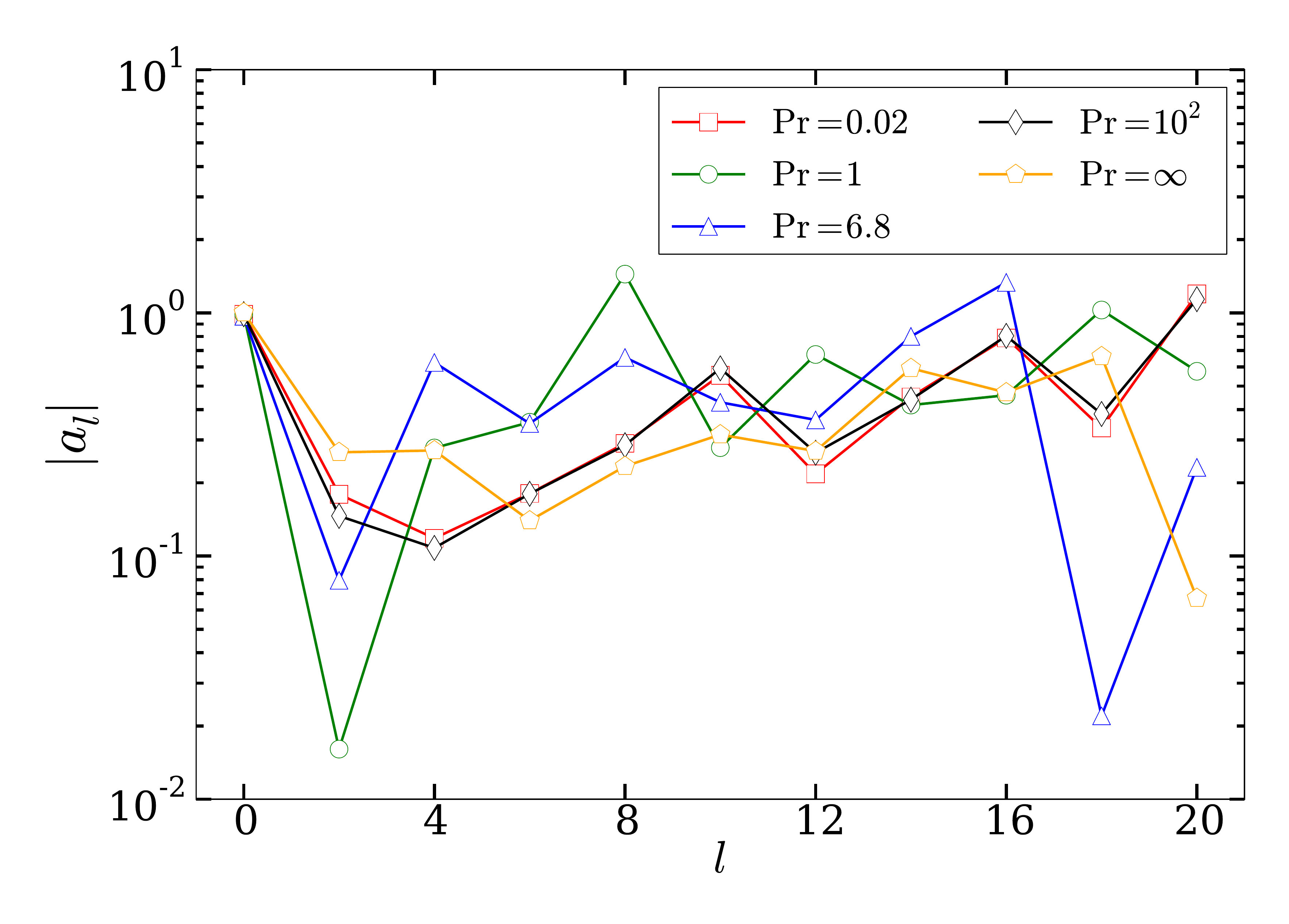}
\caption{ Plot of the absolute values of the Legendre coefficients for even $l$, $a_{l}(k)$, for various $\mathrm{Pr}$.  We choose $k$ in the inertial range: $k=50$ for $\mathrm{Pr=0.02}$, $k=40$ for $\mathrm{Pr=1}$, $k=25$ for $\mathrm{Pr=6.8}$, $k=50$ for $\mathrm{Pr=100}$, $k=65$ for $\mathrm{Pr=\infty}$.}
\label{fig:legendre} 
\end{figure}
}

In summary, the flow in RBC is weakly anisotropic with the anisotropy parameter $E_\perp/(2E_\parallel)$  bounded between 0.30 and 0.73. The anisotropy is the strongest for $\mathrm{Pr}=\infty$, and weakest for $\mathrm{Pr} \approx 1$.      In the Fourier space, the energy contents is maximum when $k_\parallel \approx k_\perp$.  

In the next section, we will describe the properties of energy transfers in RBC.

\section{Numerical results on energy transfers} \label{sec:num-ETrans} 

As discussed in Sec.~\ref{sec:theor-Etrans},  the energy transfers in RBC can be quantified using the energy flux, the shell-to-shell energy transfers, the ring-to-ring energy transfers, and the energy transfer from the parallel component of velocity to the perpendicular component of velocity.  We start with the computation of the shell-to-shell energy transfers.   

We divide the Fourier space into 18 concentric shells with their centres at ${\bf k} = (0,0,0)$.  The inner and outer radii of the $n$th shell are $k_{n-1}$ and $k_n$ respectively with $k_n$ = (0, $2$, $4$, $8$, $10.8$, $14.7$, $20.1$, $27.3$, $37.1$, $50.5$, $68.7$, $93.4$, $127.0$, $172.7$, $234.8$, $319.2$ $434.1$, $590.2$, $802.5$).   The shells in the inertial range have been binned logarithmically keeping in mind the powerlaw physics observed here.  Note that the Kolmgorov's theory of hydrodynamic turbulence predict local interactions among the logarithmically binned shells~\cite{Verma:PR2004}. The ratio $k_{n}/k_{n-1}  \approx 1.36$ has been chosen small enough   to ensure a fairly refined analysis of the shell interactions.  The first three shells  however have wider width since they have small number of modes.  In this paper, we attempt to test the locality hypothesis for RBC.  We compute the shell-to-shell energy transfers among all the shells using a technique proposed by Dar \textit{et al.}~\cite{Dar:PD2001} and Verma~\cite{Verma:PR2004}.  We perform this analysis for Prandtl numbers 0.02, 1, 6.8,  and 100.  Note that the nonlinear term ${\bf u \cdot \nabla u}$ is absent for $\mathrm{Pr}=\infty$.

\begin{figure}
\includegraphics[scale=0.63]{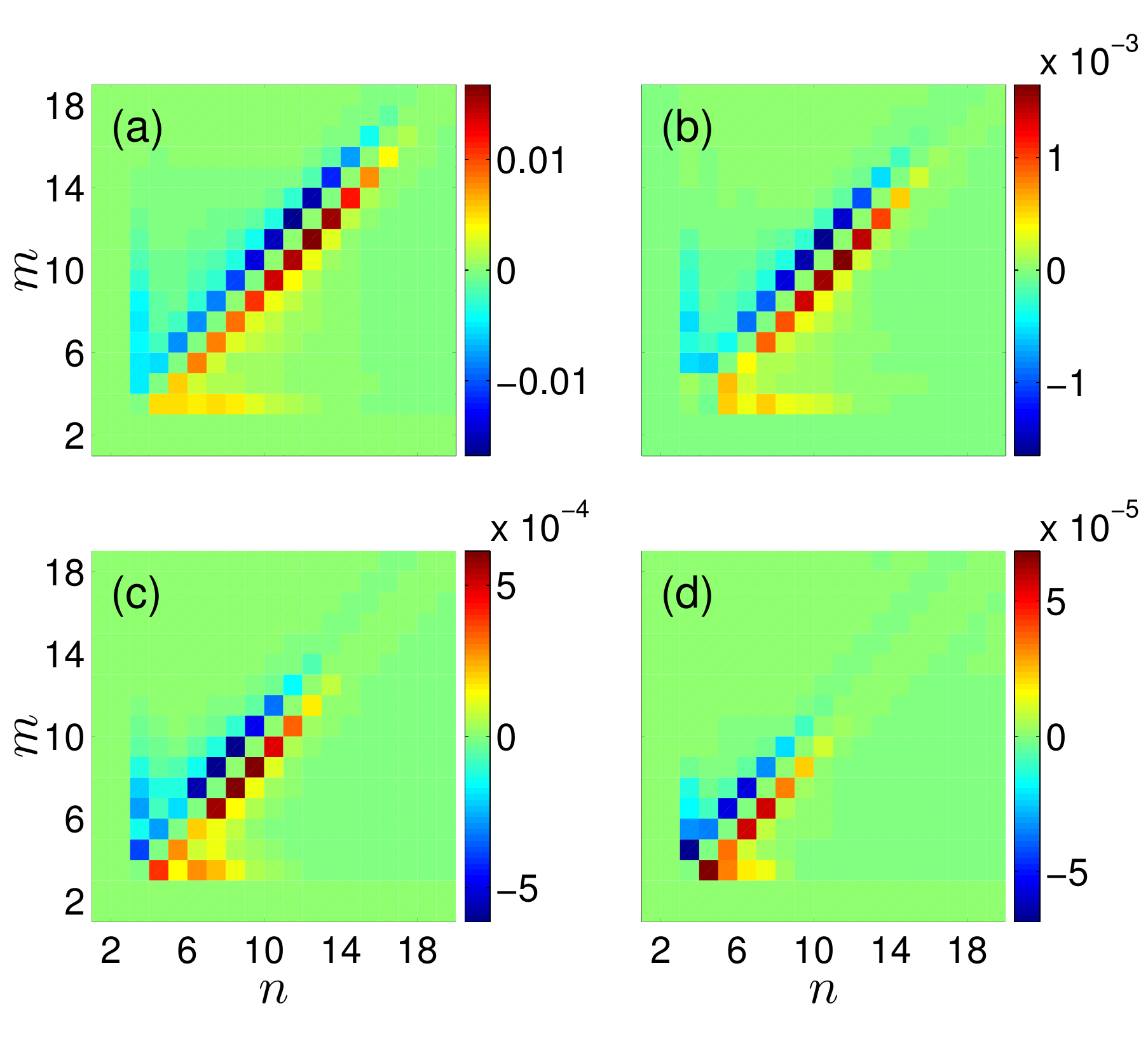}
\caption{Density plot of the shell-to-shell energy transfer rate  $T_{n}^{m}$, where $m$ of the $y$ axis   is giver shell, and $n$ of the $x$ axis is the receiver shell:  (a) $\mathrm{Pr}=0.02$, $\mathrm{Ra}=2\times10^{6}$, (b) $\mathrm{Pr}=1$, $\mathrm{Ra}=10^{8}$, (c) $\mathrm{Pr}=6.8$,  $\mathrm{Ra}=10^{8}$, and (d) $\mathrm{Pr}=100$, $\mathrm{Ra}=10^{8}$. }
\label{fig:shell2shell} 
\end{figure}

In Fig.~\ref{fig:shell2shell} we exhibit this transfer using a density plot of $T^m_n$, where $m$ along the $y$ axis is the giver shell index, and $n$ along the $x$ axis is the receiver shell index.  We observe that the most dominant energy transfer is among shells $m-1, m$ and $m+1$; the $m$th shell gives energy to $(m+1)$th shell, and receives energy from the $(m-1)$th shell.  Since the maximal energy transfer is to the neighboring shell with a larger index, we conclude that the shell-to-shell energy transfer in RBC is forward and local, very similar to that observed in hydrodynamic turbulence.  This observation is consistent with the results of Kumar \textit{et al.}~\cite{Kumar:PRE2014}  that  the turbulent RBC exhibits Kolmogorov energy spectrum ($\sim k^{-5/3}$).   We also observe a small nonlocal energy transfer from the third shell  to the shells $n=4:10$.  This is possibly due to the correlations between the energy-containing large-scales structures and the velocity field at smaller length scales.

\begin{figure}
\includegraphics[scale=0.60]{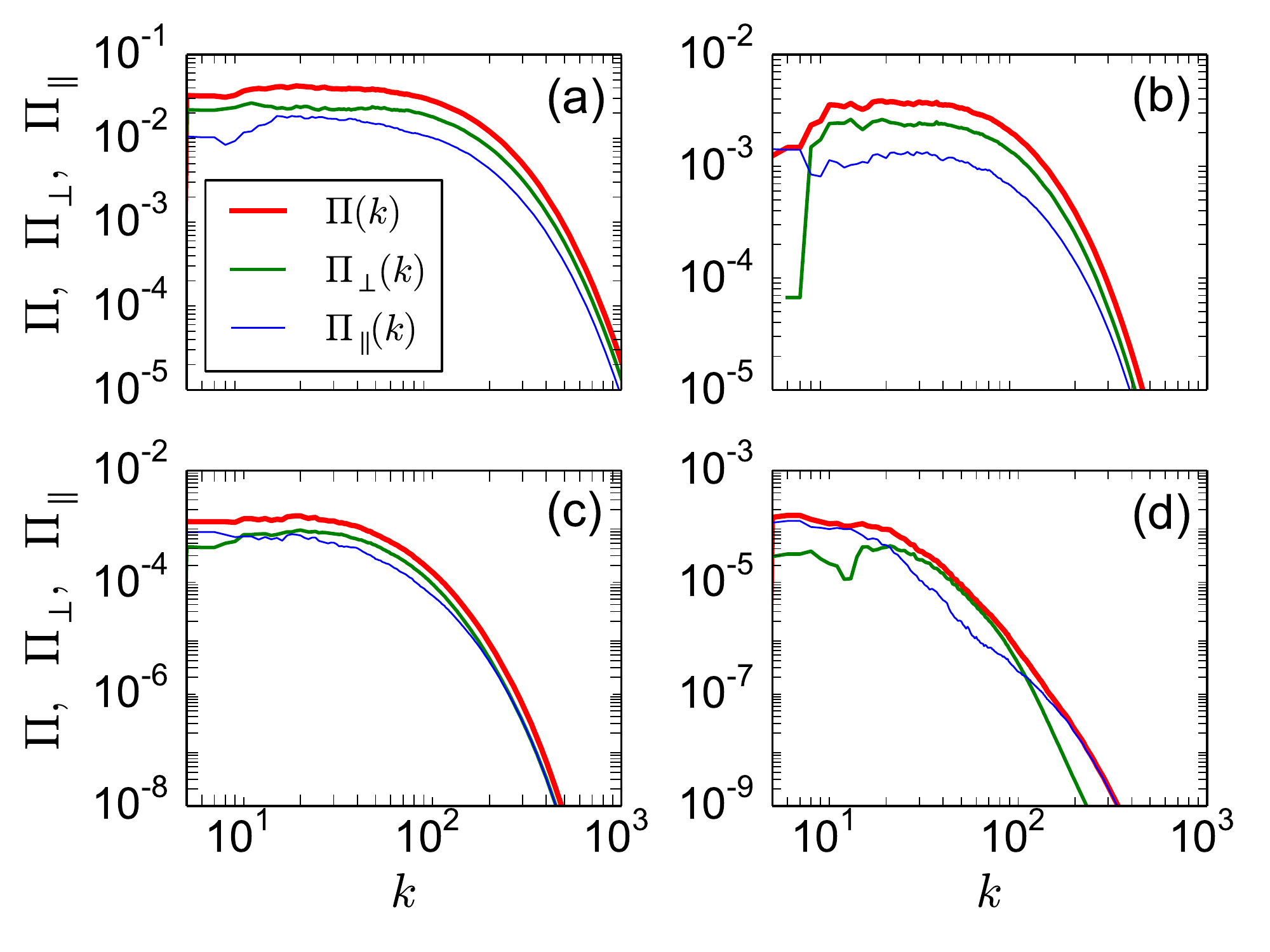}
\caption{Plots of the kinetic energy flux $\Pi\left(k\right)$, $\Pi_{\perp}\left(k\right)$, and $\Pi_{\parallel}\left(k\right)$ for (a) $\mathrm{Pr}=0.02$, $\mathrm{Ra}=2\times10^{6}$, (b) $\mathrm{Pr}=1$, $\mathrm{Ra}=10^{8}$,
(c) $\mathrm{Pr}=6.8$, $\mathrm{Ra}=10^{8}$, and (d) $\mathrm{Pr}=100$, $\mathrm{Ra}=10^{8}$. The legends for (b-d) are same as in (a).}
\label{fig:fluxes_PI_perp_para} 
\end{figure}

Next, we investigate the energy fluxes $\Pi(k), \Pi_\perp(k)$,  and $\Pi_\parallel(k)$ of RBC.  We compute the energy fluxes  for spheres with radii $1, 2, 3, ..., 1000$.  We observe positive values for all the fluxes, indicating a forward energy cascade.  We also observe a nearly constant flux in the inertial range, similar to that observed by Kumar {\em et al.}~\cite{Kumar:PRE2014}.  These observations are consistent with the forward and local shell-to-shell energy transfers discussed above.  We observe that $\Pi_\parallel(k) < \Pi_\perp(k)$ for $\mathrm{Pr} =  0.02,1$, and 6.8.  But  $\Pi_\parallel(k) > \Pi_\perp(k)$ for $\mathrm{Pr}=100$ is due to dominance of $u_\parallel$ over $u_\perp$ (see Fig.~\ref{fig:fluxes_PI_perp_para}).     Note that the fluxes vanish for $\mathrm{Pr}=\infty$ due to absence of the nonlinear term ${\bf u \cdot \nabla u}$. 

\begin{figure}
\includegraphics[scale=0.60]{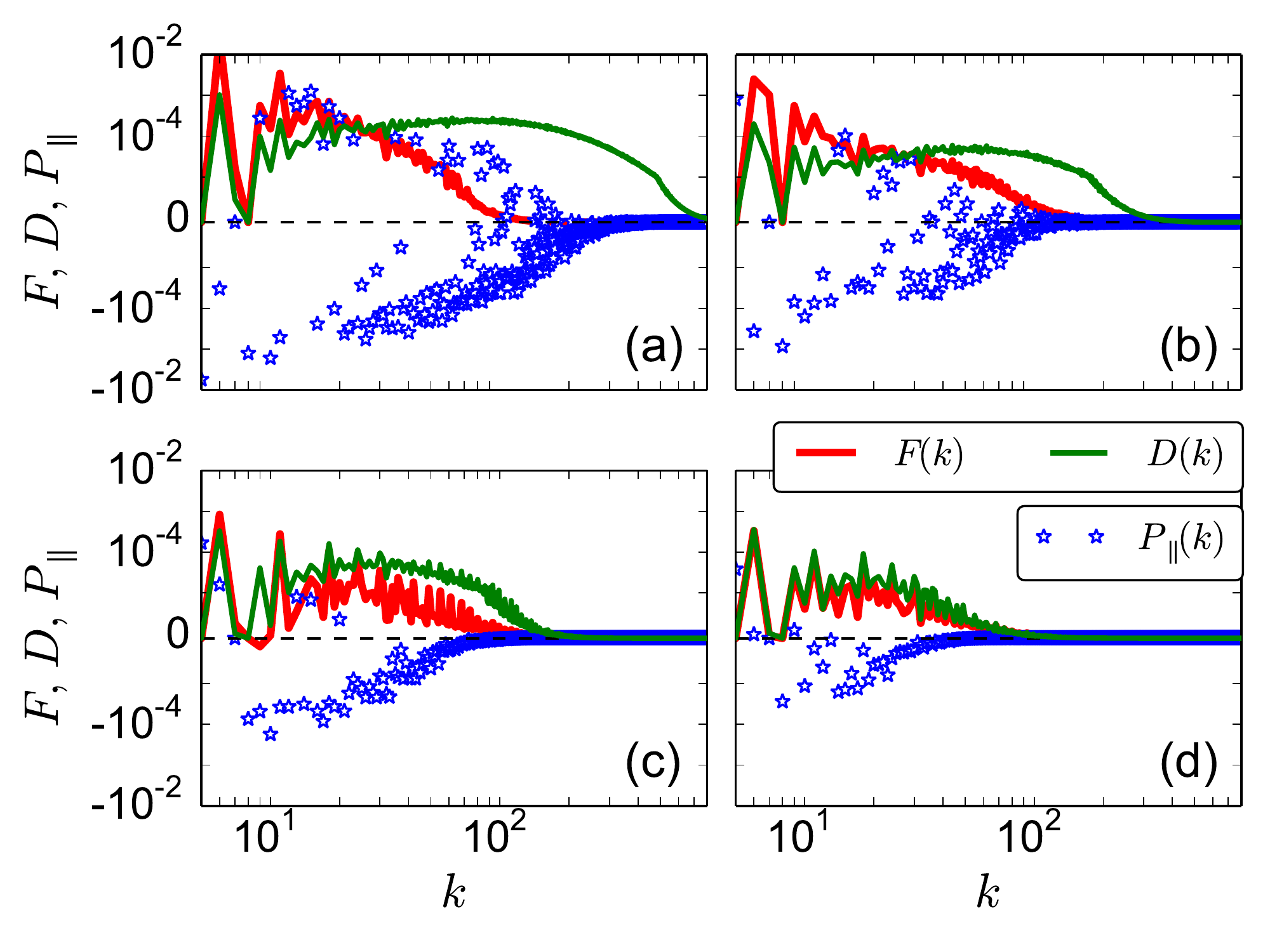}
\caption{The energy supply rate  by buoyancy $F (k)$ (red curve), viscous dissipation rate $D (k )$ (green curve), and the energy transfer rate from ${\bf u}_\perp$ to ${\bf u}_\parallel$  via pressure, $P_{\parallel}(k)$ (blue stars) for (a) $\mathrm{Pr}=0.02$, $\mathrm{Ra}=2\times10^{6}$, (b) $\mathrm{Pr}=1$, $\mathrm{Ra}=10^{8}$, (c) $\mathrm{Pr}=6.8$, $\mathrm{Ra}=10^{8}$, and (d) $\mathrm{Pr}=100$, $\mathrm{Ra}=10^{8}$. The legends for (b--d) are same as in (d).}
\label{fig:fluxes_F_D_Ppara} 
\end{figure}

As described in Sec.~\ref{sub:theor-Etrans-para-perp}, $u_\parallel(k)$ loses energy to the larger wavenumber modes due to nonlinear cascade, but also to $u_\perp(k)$ via pressure [see Eqs.~(\ref{eq:Eperp_t_eqn}-\ref{eq:Epara_t_eqn})].   In Fig.~\ref{fig:fluxes_F_D_Ppara}, we plot the energy supply rate due to buoyancy, $F(k)$, the dissipation rate, $D(k)$, and the energy transfer rate from $u_\perp(k)$ to $u_\parallel(k)$ via pressure, $P_\parallel(k)$.  As shown in  Fig.~\ref{fig:fluxes_F_D_Ppara}, $F(k) > 0$ indicating a positive energy feed to the kinetic energy by buoyancy~\cite{Kumar:PRE2014}.  Naturally $D(k) > 0$ due to viscous dissipation.  The figures shows that $P_\parallel(k) < 0$, hence $P_\perp(k) = - P_\parallel(k) > 0$, a demonstration that the energy is transferred from $u_\parallel(k)$  to $u_\perp(k)$ via pressure.  For some of the wavenumbers, $u_\parallel(k)$  receives energy from  $u_\perp(k)$ ($P_\parallel(k) >  0$), which is due to the random nature of turbulence.   Another noticeable behaviour is  relatively smaller values of $P_\parallel(k)$ for $\mathrm{Pr} = 100$, which is due to the dissipative nature of the flow (smaller $u$).

The results on energy transfers are consistent with  those of anisotropic energy distribution discussed in Sec.~\ref{sec:num-E}.  In this paper, we have not performed ring-to-ring energy transfers since they are very expensive to compute.

\section{Conclusions and Discussions} \label{sec:conclu}

In this paper we study the anisotropy in turbulent RBC for a wide range of Prandtl numbers, from very small value (0.02) to infinity.  Our conclusions are based on high-resolution numerical simulations of RBC at  high Rayleigh numbers ($\sim$ $10^6$ and $10^8$).  We quantify the anisotropy by (a) the ratio of the energies corresponding to perpendicular and parallel components of velocity, (b) ring spectrum, and (c) the energy transfers from the parallel component of the velocity to the perpendicular component.  We also compute the energy flux and shell-to-shell energy transfers.  Our results show that the flow in RBC is nearly isotropic.  Our main conclusions are as follows:
\begin{enumerate}
\item The velocity component along the buoyancy direction is stronger than  both the perpendicular components.  However, the difference is not very significant.   For $\mathrm{Pr}=1$,   $ A = E_\perp/(2 E_\parallel)$ is maximum with a value of 0.73, hence    $E_x$ and $E_y$ are approximately 70\% of $E_\parallel$.  Thus the flow is quite close to being isotropic for $\mathrm{Pr}=1$.  RBC is most anisotropic ($A \approx 0.30$) for $\mathrm{Pr}=\infty$, but here too   $E_x$ and $E_y$ are approximately 30\% of $E_\parallel$.   The aforementioned nature of anisotropy is also reflected in the energy spectra $E_\parallel(k)$ and $E_\perp(k)$.

\item Our study of ring spectrum shows that for $\mathrm{Pr}=1$, RBC is nearly isotropic.  We observe that the deviations from isotropy is strongest in the diagonal region of the Fourier space $(k_\perp \approx k_\parallel)$.  The reason for the isotropy is possibly due to an approximate cancellation of $F(k)$ (the energy feed due to  buoyancy) and $D(k)$ (the viscous dissipation).  This feature leads to a constant energy flux, as in hydrodynamic turbulence~\cite{Kumar:PRE2014}.

\item For large and infinite Prandtl number, for which anisotropy is maximal, $u_\perp$ is strong near the polar region ($\zeta \approx 0$) of the Fourier space, while $u_\parallel$ is strong near the equator ($\zeta \approx \pi/2$).  In this parameter regime, $u_\parallel$ dominates $u_\perp$, and it has a quasi-uniform structure along $z$; these results are consistent with the features  of thin plumes.

\item The reason for variation of anisotropy with Prandtl number is due to the plume structures. For large and infinite Prandtl number RBC,  the plumes are thin, and $u_{\parallel}$ is stronger than $u_{\perp}$, consistent with the earlier observations by Hansen {\em et al.} \cite{Hansen:PF1990}. The plumes tend to  spread out for lower Prandtl numbers. Still $u_{\parallel}>u_{\perp}$ due to  buoyancy. As a result $A=E_{\perp}/(2E_{\parallel})<1$ for all Prandtl numbers, but it monotonically decreases with the increase of the Prandtl number.

\item We computed the shell-to-shell energy transfer among the shells in the Fourier space.  We show that the energy transfer is local and forward, very similar to that observed for hydrodynamic turbulence (in accordance to Kolmogorov's theory).   This is consistent with the spectral studies of Kumar \textit{et al.}~\cite{Kumar:PRE2014} where they show that RBC exhibits Kolmogorov's spectrum as in hydrodynamic turbulence.  Our studies also show a small nonlocal energy transfer from the third  shell (corresponding to the large-scale convection rolls) to shells $n=4$ to 10.  The energy flux is positive and constant in the inertial range, consistent with the forward and local shell-to-shell energy transfers, a feature similar to that of hydrodynamic turbulence.

\item Buoyancy feeds energy to the parallel component of velocity, $u_\parallel$.  We show a net energy transfer from  $u_\parallel$ to $u_\perp$.  This transfer, that occurs via pressure, maintains the steady state of $u_\perp$.  
\end{enumerate}

We  remark that the anisotropy in RBC could depend on the boundary condition and geometry (box or cylinder).  Yet, the extent of anisotropy  is  expected to be approximately similar to the results presented in this paper. This is because most of the energy  of the flow resides in the bulk (away from the boundary layer).  The anisotropy studies in the boundary layer is beyond the scope of this paper.

It is also important to compare anisotropy in various buoyancy-driven flows.  In the present paper we showed that for RBC,  $E_\perp/(2E_\parallel)$ varies from 0.30 (for $\mathrm{Pr}=\infty$) to 0.73 (for $\mathrm{Pr}=1$).  Thus, RBC is nearly isotropic for the range of Prandtl numbers from 0.02 to $\infty$. The other features such as ring spectrum and shell-to-shell energy transfers also exhibit isotropy in RBC.  The anisotropy in Rayleigh-Taylor instability (RTI) varies with time during the growth of instability~\cite{Cabot:POF2013}. However in the steady state, the behaviour of RTI is quite similar to RBC, e.g., $E_\parallel$ is approximately three times that of $E_x$~\cite{Cabot:POF2013}.   Some of the anisotropy features of  Unstably Stratified Homogeneous Turbulence (USHT) are similar to that of RBC, albeit at smaller scales.   In particular, Verma {\em et al.}~\cite{Verma:IUTAM2016} showed that the Richardson number for RBC is $\mathrm{Ri}\approx16$ leading to Froude number $\mathrm{Fr}=1/\sqrt{\mathrm{Ri}}\approx0.25$, which is reasonably close to the saturated value of the Froude number of Burlot {\em et al.}~\cite{Burlot:PF2015} (see Fig.~10 of Burlot {\em et al.}~\cite{Burlot:PF2015}).

Some of the tools discussed in the paper apply to other systems with external forcing, for example,  rotating turbulence~\cite{Sagaut:book}, magnetohydrodynamic flows~\cite{Shebalin:JPP1983,Teaca:PRE2009}, quasi-static magnetohydrodynamics~\cite{Reddy:PF2014,Reddy:PP2014,Favier:PF2010a,Favier:PF2010b,Favier:JFM2011}, and rotating convection~\cite{Kunnen:PRL2008}.  Another notable and useful tool for studying anisotropy is the toroidal and poloidal decomposition of the velocity field~\cite{Sagaut:book,Favier:PF2010a,Favier:PF2010b,Favier:JFM2011}. 
These studies  provide valuable insights into such flows, but more needs to be done, specially at high resolutions. 

The present paper does not include discussion on the ring-to-ring energy transfers, as well as on the entropy spectrum ($|\theta(k)|^2$) and entropy transfers for the temperature fluctuations. These results will be presented in future.  In this paper we detail the anisotropy in the whole fluid, which is essentially dominated by the bulk flow.  Quantification of anisotropy in the boundary layer will be very useful for understanding and  modelling the boundary layer.  Thus we conclude our discussion on the anisotropy quantification in RBC.   

\begin{acknowledgments} We thank Biplab Dutta, Sandeep Reddy, Rohit Kumar,  Anando Chatterjee, and  J. K. Bhattacharjee for useful discussions and help in postprocessing. This work was supported by research grants from Indo-French Centre for the Promotion of Advanced Research (Grant No. SPO/IFCPAR/PHY/20120319) and Science and Engineering Research Board, India (Grant number SERB/F/3279/2013-14). Our numerical simulations were performed on HPC cluster of IIT Kanpur, Param Yuva at the Centre for Development of Advanced Computing (CDAC), and {\sc Shaheen} supercomputer at KAUST supercomputing laboratory, Saudi Arabia. 
\end{acknowledgments}

\end{document}